%% file: main.tex
\documentclass[sigconf]{acmart}
\usepackage{svg}
\usepackage{needspace}
\usepackage{placeins}
\usepackage{graphicx}
\usepackage{subcaption}
\usepackage{float}
\usepackage{multirow}
\usepackage{array}
\usepackage{makecell}
\usepackage{xcolor}
\usepackage{caption}
\definecolor{brownishred}{RGB}{178, 34, 34}
\definecolor{rc_note}{RGB}{186, 85, 211}

\newif\ifshowtodo
\showtodotrue

\AtBeginDocument{%
  }

\setcopyright{none}
\renewcommand\footnotetextcopyrightpermission[1]{}
\begin{document}

\title[Selection Techniques Comparison for PLV in AR]{Head, Gaze, or Finger? Comparing Object Selection Techniques in Augmented Reality for People with Low Vision}

\author{Ruijia Chen}
\affiliation{%
  \institution{University of Wisconsin-Madison}
  \city{Madison}
  \state{Wisconsin}
  \country{USA}
}
\email{ruijia.chen@wisc.edu}
\orcid{0000-0002-1655-6228}

\author{Tianyi Zhang}
\affiliation{%
  \institution{University of Wisconsin-Madison}
  \city{Madison}
  \state{Wisconsin}
  \country{USA}
}
\email{tzhang628@wisc.edu}
\orcid{0009-0007-6356-2712}

\author{Sanbrita Mondal}
\affiliation{%
  \institution{University of Wisconsin-Madison}
  \city{Madison}
  \state{Wisconsin}
  \country{USA}
}
\email{smondal4@wisc.edu}
\orcid{0000-0003-4454-8978}

\author{Yukang Yan}
\affiliation{%
  \institution{University of Rochester}
  \city{Rochester}
  \state{New York}
  \country{USA}
}
\email{yukang.yan@rochester.edu}
\orcid{0000-0001-7515-3755}

\author{Yuhang Zhao}
\affiliation{%
  \institution{University of Wisconsin-Madison}
  \city{Madison}
  \state{Wisconsin}
  \country{USA}
  }
\email{yuhang.zhao@cs.wisc.edu}
\orcid{0000-0003-3686-695X}

\renewcommand{\shortauthors}{Chen et al.}

\begin{abstract}
Augmented reality (AR) can enhance visual perception for people with low vision (PLV) by overlaying multimodal information. Selection-based augmentation further allows users to flexibly choose and augment relevant information while reducing distraction and visual clutter. However, little is known about the ability and preferences of PLV in performing object selection techniques in AR, considering their potential visual and gaze control challenges. To understand what selection techniques are suitable for PLV to support selection-based AR augmentations, we conducted a mixed-methods study with 20 PLV and 18 sighted controls who performed target selection tasks using three input techniques---head, gaze, and finger pointing with dwell-based confirmation---in two real-world scenarios (sitting vs. on the go). We found that for PLV, gaze-based selection enabled the fastest initial pointing when sitting and comparable overall selection time to head-based selection in both scenarios; however, due to reduced gaze stability, head-based selection remained the most stable and the least mentally demanding. Uniquely, participants with central vision loss preferred finger-based selection, reporting a greater sense of control. Our results provide empirical insights into accessible AR interaction techniques and selection-based vision enhancements for PLV.
\end{abstract}



\begin{teaserfigure}
    \centering
  \includegraphics[width=1\textwidth]{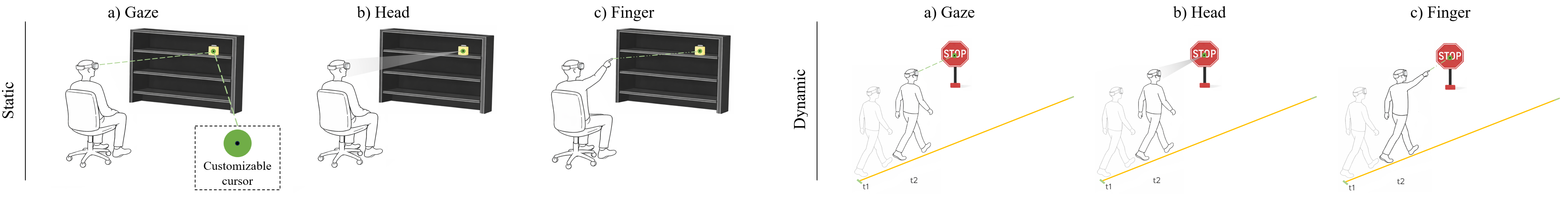}
  \caption{We investigated low-vision users' performance and experience with head-, gaze-, and finger-based selection techniques in sitting and walking scenarios.}
  \Description{}
  \label{fig:teaser}
\end{teaserfigure}

\maketitle

\fancyhead[L]{}
\fancyhead[RO]{}
\fancyhead[LE]{}
\fancyhead[R]{}
\fancyhead[LO]{Head, Gaze, or Finger? Comparing Object Selection Techniques in Augmented Reality for People with Low Vision}
\fancyhead[RE]{\shortauthors}

\input{sections/1-introduction}
\input{sections/2-related_work}
\input{sections/3-techniques}
\input{sections/4-results}
\input{sections/5-discussion}
\input{sections/6-conclusion}


\bibliographystyle{ACM-Reference-Format}
\bibliography{sections/citation}

\appendix
\input{sections/Appendix}

\end{document}

%% file: sections/1-introduction.tex
\section{Introduction}
People with low vision (PLV) have functional vision but experience uncorrectable visual impairments, such as central or peripheral vision loss, blurred vision, and night blindness \cite{NEI_LowVision}. Augmented reality (AR) technology presents potential in enhancing PLV's visual perception by overlaying multimodal information onto the physical world, such as highlighting a searched product in grocery shopping \cite{zhao2016cuesee}, distinguishing the graspable and non-graspable area of kitchen tools with AR overlays \cite{lee2024cookar}, generating virtual labels and signs to support indoor wayfinding \cite{huang2019augmented, chen2025visimark, zhao2020effectiveness}, and augmenting stairs and outlining obstacles for safe navigation \cite{fox2023using, zhao2019designing}. However, existing AR systems often assume certain types of objects to augment, overlooking users' agency in determining and selecting what they want to augment. Prior research has highlighted the importance of sense of autonomy in assistive technologies \cite{herskovitz2023hacking,pal2017agency}.

To enhance user agency, it is critical to design selection-based augmentation systems and explore how to enable PLV to efficiently and easily select objects to augment in different scenarios. While various input techniques have been designed and evaluated in AR context, such as head-, gaze-, and finger-based selections \cite{minakata2019pointing,kyto2018pinpointing,blattgerste2018advantages,lin2015investigation,bernardos2016comparison}, prior work only focuses on the performance and experiences of sighted users \cite{jalaliniya2014head,qian2017eyes,cournia2003gaze}. In contrast to sighted users, low-vision people face unique object selection challenges. For example, they may not be able to clearly perceive a small object to accurately select it, they may face difficulty controlling their gaze stably when using gaze-based selection, or they may easily lose track of the cursor when using finger-based selection. However, no research has investigated how low-vision conditions may affect users' ability, performance, and preferences in different selection techniques. 

Our research seeks to deeply understand PLV’s performance, experiences, challenges, and accessibility barriers across fundamental selection techniques, thus inspiring inclusive AR interactions and enabling selection-based augmentation systems. We ask:

\textbf{RQ1}: What are PLV's ability and performance in AR target acquisition with different selection modalities? 

\textbf{RQ2}: How  do different low-vision conditions (e.g., visual acuity and visual field) shape PLV’s performance, preferences, and experiences across selection techniques and AR scenarios?

To answer these research questions, we recruited 20 low-vision participants and 18 sighted control participants who performed target selection tasks in AR using head-, gaze-, and finger-based pointing techniques \cite{wei2025reevaluating,namnakani2023comparing, chen2021adaptive} with a dwell-based confirmation \cite{namnakani2023comparing, chen2021adaptive,wei2025reevaluating}. They conducted these tasks in both sitting and walking scenarios to represent real-world stationary and on-the-go contexts. In the stationary scenario, participants selected targets of varying sizes on a shelf. In the dynamic scenario, they walked through an indoor lab space while selecting a target located along the route.

Our findings show that, for PLV, gaze-based selection enabled the fastest initial pointing in the stationary scenario and achieved comparable overall selection time to head-based selection in both stationary and dynamic scenarios. However, PLV experienced greater difficulty in maintaining gaze dwelling stability than sighted participants. In contrast, head-based selection remained the most stable technique across target sizes, scenarios, and user groups, with significantly shorter confirmation time, fewer cursor reentries during dwelling for small target selection, and more positive perceived experience by PLV. We also identified a compounded effect of reduced acuity and restricted peripheral field on both selection efficiency and stability. Uniquely, participants with central vision loss preferred finger-based selection, reporting a greater sense of control. 

In summary, we contribute the first empirical study investigating PLV's ability and experience across typical object selection techniques in stationary and dynamic AR contexts. Our research sheds light on accessible AR interaction techniques and inspires future selection-based vision enhancements for PLV.

%% file: sections/2-related_work.tex
\section{Related Work}
We review prior work on assistive technologies for PLV in AR, as well as research on selection techniques in mixed reality.

\subsection{Assistive Technologies for PLV in AR}
Many assistive AR systems have been developed to support PLV in daily tasks, including navigation \cite{zhao2020effectiveness,chen2025visimark,fox2023using}, kitchen preparation \cite{lee2024cookar}, item searching \cite{zhao2016cuesee}, and exercise \cite{abe2025can,lee2024towards}. Some of these systems provide generic augmentations over the full field of view that enhance all visual elements \cite{hwang2014augmented,moshtael2015high,angelopoulos2019enhanced, zhao2015foresee}.
For example, Hwang and Peli \cite{hwang2014augmented} implemented an edge enhancement system on Google Glass to increase contrast for people with age-related macular degeneration.
Angelopoulos et al. \cite{angelopoulos2019enhanced} used color-coded wireframes to present depth information for people with retinitis pigmentosa. 
However, these systems indiscriminately enhance all visual information without considering what users want to augment, potentially resulting in visual clutter and distraction. %

Some other systems instead focus on augmenting predefined object categories for certain contexts \cite{lee2024cookar,zhao2016cuesee,chen2025visimark,abe2025can, fox2023using, appleMagnifyDescribe, huang2019augmented}. For example, Huang et al.~\cite{huang2019augmented} designed an AR system that detects text signs and replaces them with enhanced text or verbal descriptions to assist sign reading. CookAR~\cite{lee2024cookar} highlights kitchen tool affordances (e.g., knife handles vs. blade) to support safe kitchen preparation.
VisiMark~\cite{chen2025visimark} augments selected landmark types in AR to support indoor navigation. 
While these systems constrain augmentation to relevant object types, they do not allow users to proactively select what to augment. Since user needs vary across tasks, environments, and personal preferences \cite{kane2009freedom}, it is critical to support user agency in flexibly specifying augmentation targets \cite{herskovitz2023hacking}. However, no work has investigated how to enable PLV users to easily and accurately select objects in real-world AR environments.

The most relevant line of work explores region-based visual enhancement that follows the user's finger or gaze ~\cite{kaya2024virtual,stearns2018design,maus2020gaze,masnadi2020vriassist}. For example, Stearns et al.~\cite{stearns2017augmented,stearns2018design} designed an AR magnification system that combines a finger-worn camera or handheld phone with HoloLens to present magnified content. It allows users to control the magnified area by moving their finger or the phone. 
Maus et al. \cite{maus2020gaze} designed a screen magnifier that follows users' gaze movement. 
Similarly, VRiAssist \cite{masnadi2020vriassist} developed gaze-following visual corrections to  compensate for the distorted area of the user’s eyes in virtual reality. %
However, these systems often directly attach enhancements to finger-pointed or gaze position and arbitrarily augment all regions that the user bypasses beyond their intent, causing the ``Midas touch'' issue \cite{jacob1990you} for augmentations. 

GazePrompt~\cite{wang2024gazeprompt} demonstrated the potential of intent-aware augmentations by detecting PLV's gaze behaviors and generating augmentations on a certain word or line in reading tasks. However, it detected user intent implicitly and did not allow direct control from the users, limiting their sense of autonomy. Our work emphasizes the importance of user-initiated augmentations, investigating suitable AR selection techniques for PLV to enable them to flexibly select objects they want to augment in various daily contexts. 

\subsection{Selection Techniques in Mixed Reality}

Using different body parts as input for selection in mixed reality has long been studied for sighted users.
For example, head-based target selection \cite{minakata2019pointing,kyto2018pinpointing,blattgerste2018advantages} allows users to aim at targets via head orientation and are widely supported in commercial mixed reality headsets such as Meta Quest~\cite{metaHead}, Microsoft HoloLens~\cite{microsoftHeadgazeCommit}, and Apple Vision Pro~\cite{applePointer}.
With the development of hand tracking technologies, direct hand interactions have become a popular technique that enables mid-air pointing and selection without holding controllers \cite{lin2015investigation,bernardos2016comparison,gavgiotaki2023gesture,whiffing2026understanding,nguyen2023hand}. Gaze-based selection \cite{minakata2019pointing,kyto2018pinpointing,blattgerste2018advantages} relies on eye tracking technology to infer users’ visual focus and identify the target object.
To mitigate unintended target activation during pointing, prior research has also explored different confirmation mechanisms, such as dwell time \cite{namnakani2023comparing,li2025effects,paulus2021usability,xu2019pointing} and gestures \cite{mohan2018dualgaze,bace2016ubigaze,fashimpaur2020pinchtype,mutasim2021pinch}.
Beyond conventional ``pointing and confirming'' selection, alternative selection techniques have been designed to improve efficiency, such as motion matching selection \cite{velloso2017motion,drewes2019dialplates,esteves2015orbits,esteves2020comparing}.

Previous studies have compared various techniques and shown advantages and drawbacks of different techniques. For example, Blattgerste et al. \cite{blattgerste2018advantages} found that aiming targets via gaze was significantly faster and less exhausting than via head. Similarly, Jalaliniya et al. \cite{jalaliniya2014head} found that gaze pointing was significantly faster than head pointing, but head motion was more accurate and convenient. In contrast, Qian et al. \cite{qian2017eyes} found that gaze-based selection offered worse performance in terms of error rate, selection time, and throughput compared to head-based selection. While head-based pointing offers hands-free and stable interactions, prior work suggests that finger-based pointing achieves higher perceived usability, with no significant differences in task completion time \cite{bernardos2016comparison}.

Some research has explored how to combine different inputs to facilitate target selection, such as gaze and finger \cite{wagner2023fitts,lystbaek2022gaze,sidenmark2024cone,kim2025pinchcatcher,jeong2023gazehand}, gaze and head \cite{kyto2018pinpointing,sidenmark2024cone}, and head and finger \cite{kyto2018pinpointing,sidenmark2024cone,jeong2023gazehand}. For example, Kytö et al. \cite{kyto2018pinpointing} decomposed selection into two stages---pointing and refinement---and investigated combinations of head, gaze, and finger input across these stages for precise selection. 
Sidenmark et al. \cite{sidenmark2024cone} proposed Cone\&Bubble, a two-stage technique in which gaze supported rapid candidate selection and controller input enabled precise adjustment.
Other works explored gaze-hand coordination for region selection \cite{shi2023exploring}, target selection at different depths using gaze with a controller \cite{chen2023gazeraycursor} or with hands \cite{wagner2024gaze}, and multi-object selection through gestures assisted by gaze \cite{kim2025pinchcatcher}.

While prior research has investigated how basic selection techniques can be used individually or in combination to improve target selection, they have primarily focused on sighted users. To our knowledge, no work has examined how PLV perceive and compare these basic selection techniques. Due to their diverse visual conditions, PLV may face distinct challenges and perform differently with these selection techniques. We aim to fill this gap by evaluating the performance and experiences of PLV using head-, gaze-, and finger-based selection techniques in comparison to sighted users.

%% file: sections/3-techniques.tex
\section{Methods}
Our goal is to investigate three commonly used AR selection techniques (head-, gaze-, and finger-based \cite{minakata2019pointing,kyto2018pinpointing,blattgerste2018advantages,bernardos2016comparison}) with PLV, understanding their experiences, challenges, and interaction patterns in relation to visual abilities. 
We collected low-vision and sighted participants’ performance, behavioral, and experiential data in both stationary and dynamic scenarios and analyzed the results using qualitative and quantitative methods. This study was approved by the Institutional Review Board (IRB) at our university.

\subsection{Participants}\label{subsec:method_participants}
We recruited 20 PLV (L1-L20) and 18 sighted control participants (S1-S18) for our study. Our low-ision participants included 8 females, 11 males, and one non-binary, whose ages range from 19 to 91 (Mean = 58.7, SD = 19.6). Participants covered a wide range of low-vision conditions with a relatively balanced distribution across low vs. high visual acuity, and mild vs. severe peripheral field loss (see Table \ref{tab:demographics_plv} in Appendix \ref{sec:demographics_plv}). 
We recruited low-vision participants from a local low-vision clinic and via our university research email service. 
Two participants did not complete all tasks: L19 did not complete tasks in the dynamic scenario because of time constraints, and L20 did not complete small target selection tasks in the dynamic scenario due to low visibility. Participants were compensated at \$20/hour and reimbursed up to \$30 for travel expenses.

Our sighted participants included 11 females and 7 males, aged from 20 to 73 (Mean = 36.1, SD = 15.6; see Table \ref{sec:demographics_sighted} in Appendix \ref{sec:demographics_sighted}). They were recruited via our university research email service. All sighted participants had visual acuity of at least 20/25 in the better eye (20/20 with correction) and intact visual fields. They were compensated at \$10/hour.
PLV received higher compensation because of increased participation efforts.

\subsection{Three selection techniques: Head-, Gaze-, and Finger-based Selection} \label{subsec:Three selection techniques: Head-, Gaze-, and Finger-based}
 Object selection includes two stages: (1) \textit{Pointing:} the user specifies a target through pointing, and (2) \textit{Confirmation:} the user triggers the selection through certain interaction techniques \cite{blattgerste2018advantages}. We evaluated three pointing techniques: (1) head-based pointing \cite{minakata2019pointing,kyto2018pinpointing,blattgerste2018advantages}, where users aim at a target by moving their head and a cursor locked to the center of the AR display moves along\footnote{We adjusted the cursor position on the AR display for some low-vision participants with central vision loss to ensure the visibility of the cursor.}; 
(2) gaze-based pointing \cite{minakata2019pointing,kyto2018pinpointing,blattgerste2018advantages}, where users aim at a target via eye movement and the cursor follows the gaze position; and (3) finger-based pointing \cite{lin2015investigation,bernardos2016comparison,gavgiotaki2023gesture,whiffing2026understanding}, where users point at a target using their index finger and the cursor follows a ray cast from the finger.

For selection confirmation, we adopted a commonly used dwell-based method. Prior work has shown that dwell-based confirmation is effective when hands-free interaction and high reliability are required \cite{wei2025reevaluating}. The dwell time threshold was set to 0.8s based on prior literature \cite{namnakani2023comparing, chen2021adaptive} and tested in a non-representative pilot study with two sighted users and four low-vision users. A visible cursor was presented to indicate the current pointing location in the 3D space, following the prior finding that visual cursor feedback improves dwell-based selection performance \cite{wei2025reevaluating}. The cursor was
a circle with a radius of 1.5$^\circ$ visual angle at 1 meter. The cursor size was chosen to align with previously reported eye tracker accuracy (1.3–1.8°)~\cite{baumann2023neon} to reduce the impact of eye tracking offset on selection accuracy. 
The dwell timer was triggered when the cursor overlapped with a target. Once the dwell time reached 0.8s, the system played an auditory cue (``Ding'') to indicate task completion; otherwise, the trial timed out after 20s.

\subsection{System Implementation}
To evaluate the AR selection techniques in a real-world setting, we implemented an end-to-end pipeline including a head-mounted interface that supports the target selection techniques and a backend for real-world object segmentation. The frontend ran on a Meta Quest 3 headset with an integrated Pupil Labs Neon eye tracker (60 Hz, synchronized with head and finger tracking) \cite{pupillabsNeonOverview}, handling user input sensing and feedback rendering. The backend ran on a laptop with an NVIDIA GeForce RTX 4080 GPU, performing real-time target detection through computer vision models.

We ran a real-time object detection model to identify task targets (jars and stop signs in our study; see Section~\ref{subsec:environment_setup}). The detection model was based on YOLOv8~\cite{yolov8_ultralytics} and fine-tuned with a combination of MS-COCO images containing jars and glasses~\cite{lin2015mscoco}, stop sign images from the Mapillary Vistas dataset~\cite{neuhold2017mapillary}, and additional images captured in the physical lab environment used for the study.

While the target identification mainly relied on computer vision, missed detection due to motion blur or lighting changes in the real world could affect selection performance. We thus implemented a fallback mechanism by tracking the target position through a spatial anchor (i.e., persistent spatial references in the AR coordinates to represent the positions and orientations of a virtual object in the physical world \cite{muller2017remote,meta_unity_spatial_anchors_best_practices}). We created a spatial anchor for each target at the beginning of each selection trial using the detection result from the starting frame. When the vision model failed to recognize a target in a given frame, the system fell back to the spatial anchor. Because spatial anchors may drift as the user moves \cite{meta_unity_spatial_anchors_best_practices}, especially in the dynamic scenario, we updated the spatial anchor manually when needed (e.g., the target was correctly recognized but the spatial anchor drifted) based on the latest target detection result.

\subsection{Gaze Calibration and Cursor Adjustment}\label{subsec:gaze_cursor_adjustment}
To ensure high quality gaze tracking for PLV, we took additional steps for gaze calibration. Our Neon eye tracker adopted machine learning-based gaze estimation without traditional calibration \cite{pupillabsNeonOverview}. Based on the initial gaze estimation, Neon allows a simple post-hoc offset correction to compensate for individual differences: the user stares at a target in the center field while a researcher can manually drag the estimated cursor position to the target \cite{pupillabsNeonUsing}. However, this method does not consider different offsets across different viewing directions, which may cause more severe issues for PLV. We thus expanded this post-hoc offset correction by considering different viewing directions through a customizable interface.

Our calibration interface rendered 16 target dots in a $4 \times 4$ layout. All targets were world-anchored, allowing participants to move their heads freely to fixate on each target. This 3D setup can better reflect real-world viewing behaviors than screen-based calibration.
Participants were asked to fixate on each calibration target while a local offset was measured (i.e., the angular difference between the target position and the estimated gaze direction). We then applied these local offsets in real-time gaze estimation. We had two offset correction methods: (1) an average global offset across all local offsets, or (2) an average local offset derived from the four calibration targets nearest to the current gaze estimate in angular space, inspired by k-nearest neighbors algorithm
\cite{kerr2019real,jiang2019appearance}. We validated both methods with a 5-dot validation interface for each participant and applied the method that had smaller angular errors. 

To ensure calibration accessibility, we allowed target size and color customization, verbally instructed target location if it fell out of users' functional field, and adopted dominant-eye-based data collection if a user exhibited inconsistent gaze behaviors between two eyes or had one untrackable eye, following prior work \cite{wang2023understanding}.  

Our calibration resulted in a mean angular error of 1.51° (SD = 0.58) for 15 PLV (excluding L02, whose validation data were accidentally deleted, and four participants with central scotomas), and 1.45° (SD = 0.57) for 18 sighted participants. Both values fall within the reported eye tracker accuracy range (1.3–1.8°)~\cite{baumann2023neon}.

Notably, four participants experienced central vision loss (CVL) with central scotomas (i.e., central dark spots). They faced difficulty with calibration as they were unable to see the calibration targets with central vision. Instead, they relied on the \textit{preferred retinal locus} (PRL, a "pseudo-fovea" in the peripheral retinal region close to central vision) to view visual content \cite{whittaker1988eccentric, crossland2005preferred}. However, because people with CVL often rely on multiple PRLs \cite{crossland2005preferred,lei1997using}, our current offset correction method for gaze tracking does not apply.
We thus manually adjusted the gaze cursor to one of the user's PRLs to ensure cursor visibility, enabling them to control the cursor via eye movement (although the cursor does not always align with their preferred PRL when viewing a certain area). We discuss their unique experiences in Section \ref{subsubsec:CVL}.

To reduce eye tracking instability and signal loss caused by blinking, we applied a histogram-based filter for gaze tracking~\cite{blattgerste2018advantages} with a 10 frames filter window determined by a pilot study with both sighted and low-vision participants.

\subsection{Environments: Two Real-World Scenarios} \label{subsec:environment_setup}
We asked participants to conduct real object selection in AR instead of virtual target selection to better reflect real-world experiences, as virtual lighting and artificially defined target boundaries may alter PLV’s perception. We included two scenarios---selecting objects on a shelf and selecting targets while walking---to represent typical real-world stationary and dynamic contexts. Since target sizes may also significantly influence PLV's viewing and selection ability, we introduced two target sizes in each scenario.

\begin{figure}[]
    \centering
    \begin{subfigure}{0.23\textwidth}
        \centering
    \includegraphics[width=\textwidth]{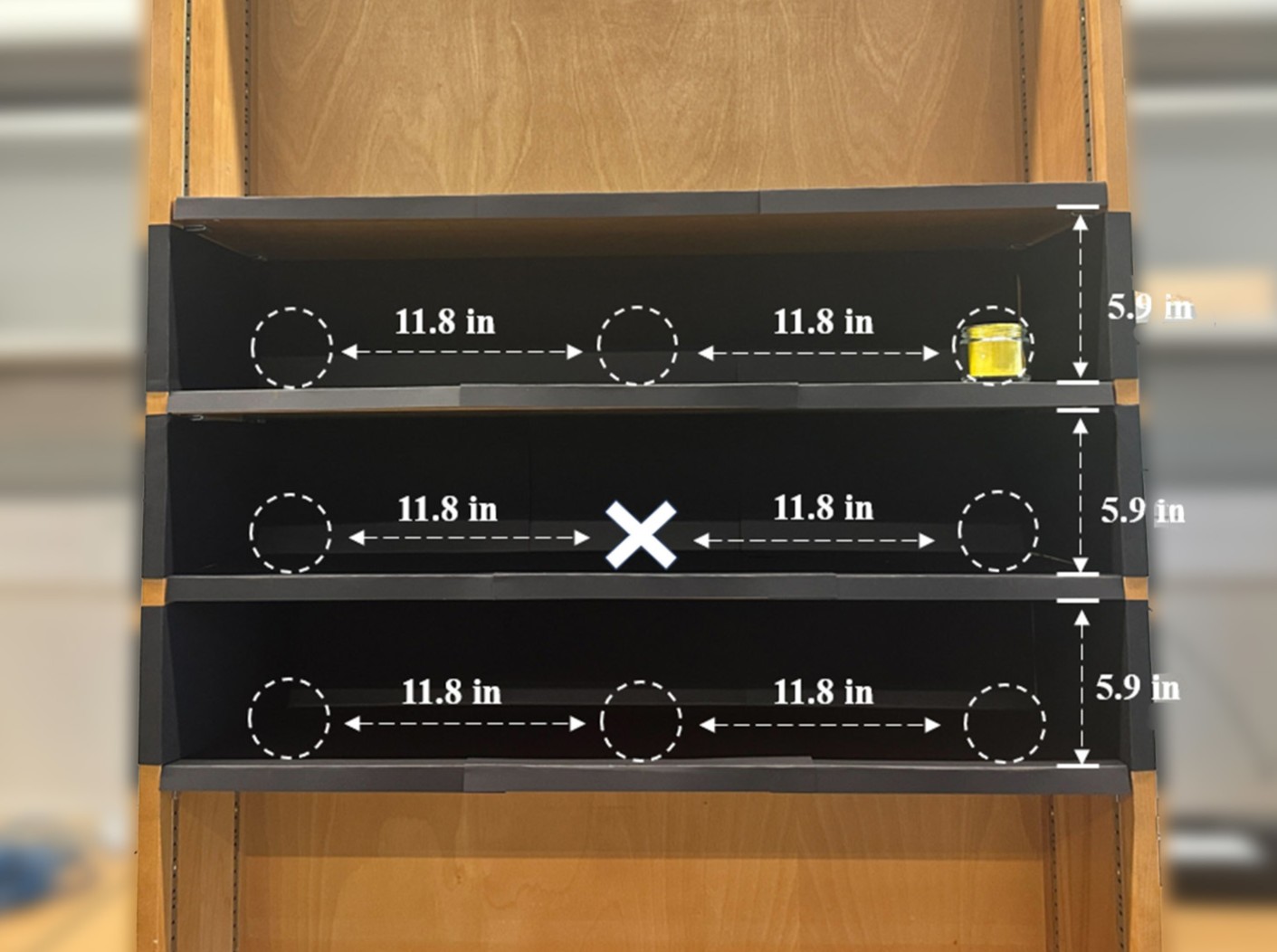}
    \end{subfigure}
    \begin{subfigure}{0.23\textwidth}
        \centering
    \includegraphics[width=\textwidth]{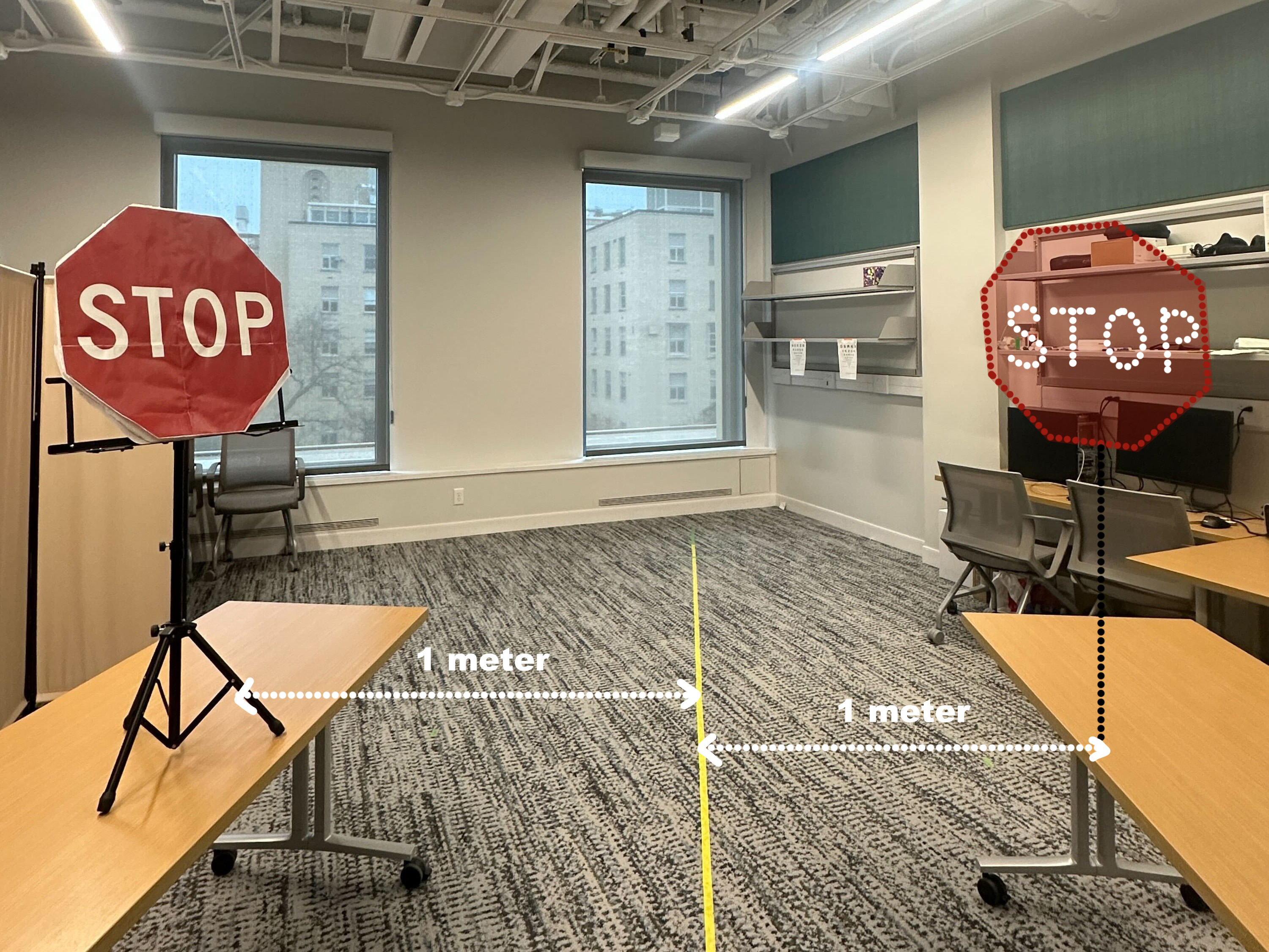}
    \end{subfigure}
    \caption{Stationary scenario setup (left): a shelf with eight target locations (3 $\times$ 3 grid, center excluded). Dynamic scenario setup (right): an 8m walking route with a stop sign 1 m from the path.}
    \Description{Left: Static task layout: a three-row shelf with eight possible target locations arranged in a 3 $\times$ 3 grid (center excluded).
    Right: Dynamic task setup: an 8 m walking route with a stop sign placed 1m away on either side of the path.}
    \label{fig:labsetup}
    \vspace{-3ex}
\end{figure}

We conducted the study in a well-lit indoor lab environment. For the stationary scenario, participants sat one meter in front of a shelf with the shelf center aligned with their eye level (Figure~\ref{fig:labsetup} left). 
We prepared target objects in two common household container sizes (2 oz and 24 oz). 
The targets were yellow in high contrast against the black background. The shelf was arranged in a $3 \times 3$ layout, providing eight target directions and positions (center excluded).

For the dynamic scenario, we used two stop signs as targets, which are commonly encountered in outdoor navigation environments, measuring 6 and 18 inches in width, respectively. 
The target sizes were determined based on pedestrian signal specifications from the Manual on Uniform Traffic Control Devices \cite{FHWA_MUTCD_Part4E_2003} and typical bus stop sign dimensions. We designed a walking route of 8 meters and marked the route on the floor using colored tape to indicate walking direction. At the midpoint of the route, we defined two possible target locations, each positioned one meter away from the walking path on opposite sides of the route, to avoid bias toward a single viewing direction (Figure~\ref{fig:labsetup} right). The target height aligned with the participants' eye level.

\subsection{Procedure} \label{subsec:procedure}
The study consisted of a single session (1.5–2 hours for sighted participants; 2–2.5 hours for PLV). We started with an initial interview to collect participants’ demographic information, visual conditions, and prior experience with AR.

\textbf{\textit{Visual ability test.}} For PVL, we then conducted a visual ability test. We first measured participants’ visual acuity with wall-mounted letter-size ETDRS 1 and ETDRS 2 logMAR charts \cite{ferris1982new} for the right and left eye, respectively.
Our visual acuity test covered a range from 20/10 to 20/400. 
We also measured participants’ visual field with a simplified visual field test interface based on prior work \cite{wang2023understanding}, covering $60 ^\circ \times 44^\circ$ central visual field.
For participants whose visual acuity or visual field fell outside the measurable range of our tests, we reported their self-reported values.

\textbf{\textit{Selection in Stationary Scenario.}} 
After eye tracking calibration, participants completed target selection tasks in a stationary scenario in front of a shelf using three selection techniques, with 16 trials per technique (2 target sizes $\times$ 8 target locations). Before each trial, the environment was temporarily blacked out, and participants fixated on a virtual white crosshair anchored at the center of the shelf to ensure a consistent starting point across trials. Once an audio cue was played that indicated the start of a trail, the scene became visible. Participants were asked to select the target on the shelf as quickly and accurately as possible. A researcher manually adjusted the target position on the shelf between trials when the AR display was blacked out. A trial ended with an auditory cue upon selection or timeout. After selection trials with each technique, participants rated technique effectiveness and task workload through NASA-TLX \cite{hart1988development} on 7-point Likert scales. We counterbalanced the order of techniques and target sizes using Latin Square \cite{richardson2018use} and randomized the order of target position.

\textbf{\textit{Selection in Dynamic Scenario.}} 
Participants then completed selection tasks in a dynamic scenario by walking along a predefined route while selecting a target, with four trials per technique (2 target sizes $\times$ 2 target locations).
Participants were asked to walk at a comfortable, self-determined pace (i.e., without running) to reflect their real-world walking speed and to complete the target selection as quickly as possible. They were encouraged to avoid unnecessary slowing or stopping, but were allowed to pause if it was necessary to complete the selection. Participants needed to complete the selection task (or timeout) and continued walking to pass the end marker of the route to end the trial. A similar target repositioning method to the stationary scenario was used between trials. After each technique, participants rated technique effectiveness and task workload again. Counterbalancing and randomization strategies mirrored the stationary scenario.

We ended our study with a semi-structured interview in which participants ranked the three selection techniques and discussed their selection experiences, challenges, and suggestions.

\subsection{Analysis}
We analyzed the quantitative data from 18 low-vision participants (L01–L18) and 18 sighted participants (S01-S18) as two low-vision participants (L19, L20) did not complete all selection tasks (Section \ref{subsec:method_participants}). Our qualitative analysis involved all 20 low-vision participants.

\subsubsection{Quantitative Analysis}\label{subsubsec:quan_analysis} We quantitatively analyzed the impact of technique and visual condition on target selection performance. 

\textbf{Measures.}
We divide the target selection process into a pointing stage and a confirmation stage (Section \ref{subsec:Three selection techniques: Head-, Gaze-, and Finger-based}). We define the time point that splits these two stages as the time of \textit{First Attempt}, which represents the moment when a participant indicates their first attempt to dwell on a target. This concept distinguishes the ``true'' first landing from the time when the cursor first overlaps with the target to mitigate impact of potential ``Midas Touch'' \cite{jacob1990you}. For gaze-based selection, the First Attempt moment is detected as the first fixation landing on the target; and for head- or finger-based selection, it is detected when the cursor remains overlapping with the target for at least 70 ms (the minimum fixation durations ~\cite{PupilLabsFixationDetector2024}). Based on the First Attempt concept, we define performance measures below.

For the static task, we defined four measures: (1) \textit{SelectionTime} is the time spanning from the selection task onset to the moment when the dwell-time threshold was triggered or timed out. It measures the whole selection efficiency. (2) \textit{PointingTime} is the time from the task onset to First Attempt moment. It measures the pointing efficiency. (3) \textit{ConfirmationTime} is the time from the First Attempt moment to the selection completion or timeout. Note that participants may fail to maintain the cursor in the target for the dwell time threshold and try again before timeout. (4) \textit{ReentryTimes} is the number of times the cursor reentered the target boundary after the First Attempt. Both \textit{ConfirmationTime} and \textit{ReentryTimes} indicate a user's ability to maintain cursor stability within a target.

For the dynamic task, we defined two additional measures for walking performance beyond the above measures: (1) \textit{WalkingTime}, the time from the onset of the walking task until the participant crossed the endpoint; and (2) \textit{SelectionTimeInWalking}, defined as the ratio of \textit{SelectionTime} to \textit{WalkingTime}, reflecting the interference between the selection and walking tasks in this dual-task setting.

All participants completed the selection tasks within the predefined time limit, except for L14 in the dynamic task. L14 experienced timeouts in 1/4, 2/4, and 3/4 trials with head-, gaze-, and finger-based selection, respectively. For these trials, we set \textit{SelectionTime} to the timeout limit (20~s) and \textit{ReentryTimes} to the maximum observed across PLV participants (12 times). 

\textbf{Factors \& Tests.} \label{subsubsec:method_compareSighted&PLV}
We have two within-subject factors: \textit{Technique} (three levels: head- vs. gaze- vs. finger-based) and \textit{Size} (two levels: large vs. small) and a between-subject factor \textit{Vision} (two levels: sighted vs. low-vision). To validate the counterbalancing, we involved another within-subject factor \textit{Order} and found no significant effect of \textit{Order} on any measures.

We checked the normality of each measure using the Shapiro-Wilk test. Since none of these measures were normally distributed, we applied the Aligned Rank Transform (ART) ANOVA \cite{wobbrock2011aligned} to evaluate the effect of different factors on different measures. We conducted post-hoc comparisons using Tukey’s HSD correction if significant effects were observed. We report effect sizes using partial eta squared ($\eta ^2_p$), with 0.01, 0.06, 0.14 representing the thresholds of small, medium and large effects \cite{cohen2013statistical}.

For PLV, we further examined the impact of low-vision conditions (reduced visual acuity, reduced peripheral vision) on user performance. We included two additional between-subject factors: \textit{PeripheralVision} and \textit{VisualAcuity}. We defined \textit{PeripheralVision} into two levels: \textit{SevereFieldLoss}, defined as moderate to severe peripheral vision loss with a remaining visual field of less than 60°, corresponding to the far peripheral vision threshold~\cite{simpson2017mini}, and \textit{MildFieldLoss}, defined as mild to no peripheral vision loss with a visual field wider than 60°. We thus have seven participants in the \textit{SevereFieldLoss} level and eleven in the \textit{MildFieldLoss} level. \textit{VisualAcuity} also has two levels, \textit{LowAcuity} and \textit{HighAcuity}, using 20/100 in the better eye as the threshold \cite{afbVisionLegal}. This resulted in nine participants in the \textit{LowAcuity} level and nine in the \textit{HighAcuity} level.

We adopted mixed-effects modeling for repeated measures~\cite{galecki2012linear,breslow1993approximate,tsandilas2024illusory}
to handle the unbalanced sample sizes across conditions. Specifically, we used linear mixed-effects models (LMMs) \cite{galecki2012linear} for \textit{SelectionTime}, \textit{PointingTime}, and \textit{ConfirmationTime}, after applying a log transformation to positively skewed variables to improve normality and variance stability. For \textit{ReentryTimes}, we used generalized linear mixed-effects models (GLMMs) \cite{breslow1993approximate} with a negative binomial distribution. All models included participant as a random intercept, and we conducted post-hoc comparisons using Tukey’s HSD correction when significant effects were observed. No order effects were found for any measures.

\subsubsection{Qualitative Analysis}
We audio-recorded all study sessions, transcribed them using Whisper \cite{radford2022whisper}, and manually fixed transcription errors. We analyzed the transcripts using thematic analysis \cite{braun2006using,clarke2017thematic}. Two researchers independently open-coded the same 10 transcripts (5 PLV, 5 sighted; 26.3\% of the data) and collaboratively developed an initial codebook by resolving disagreements through discussion. The remaining transcripts were then coded independently, with periodic checks to ensure consistency. The codebook was iteratively refined.

We developed themes from the codes using a combination of inductive and deductive approaches \cite{braun2006using}. Guided by our research goal of understanding different low-vision users' experiences and challenges with AR selection techniques, we identified high-level themes deductively and generated sub-themes inductively by clustering relevant codes using axial coding and affinity diagrams. Researchers cross-referenced the data and codebook to iteratively refine themes and ensure appropriate code categorization.

%% file: sections/4-results.tex
\section{Results}
To answer our two research questions, we first compare the selection performance and behaviors between low-vision and sighted participants through both quantitative and qualitative results. We then examine how visual abilities (visual acuity and visual field loss) affect selection performance among PLV. Full statistical results are summarized in Table~\ref{tab:stats_sightedandPLV} and \ref{tab:stats_va_fov} in Appendix \ref{sec:stats_plv_sighted} and \ref{sec:stats_va_fov}.

\subsection{Selection Performance of Low-Vision Users}
We report low-vision participants' target selection efficiency, stability, and impact on walking performance across different selection techniques, in comparison to sighted controls to reflect their target selection ability and preferences.

\begin{figure*}[]
    \centering
    \begin{subfigure}{0.49\textwidth}
        \centering
    \includegraphics[width=\textwidth]{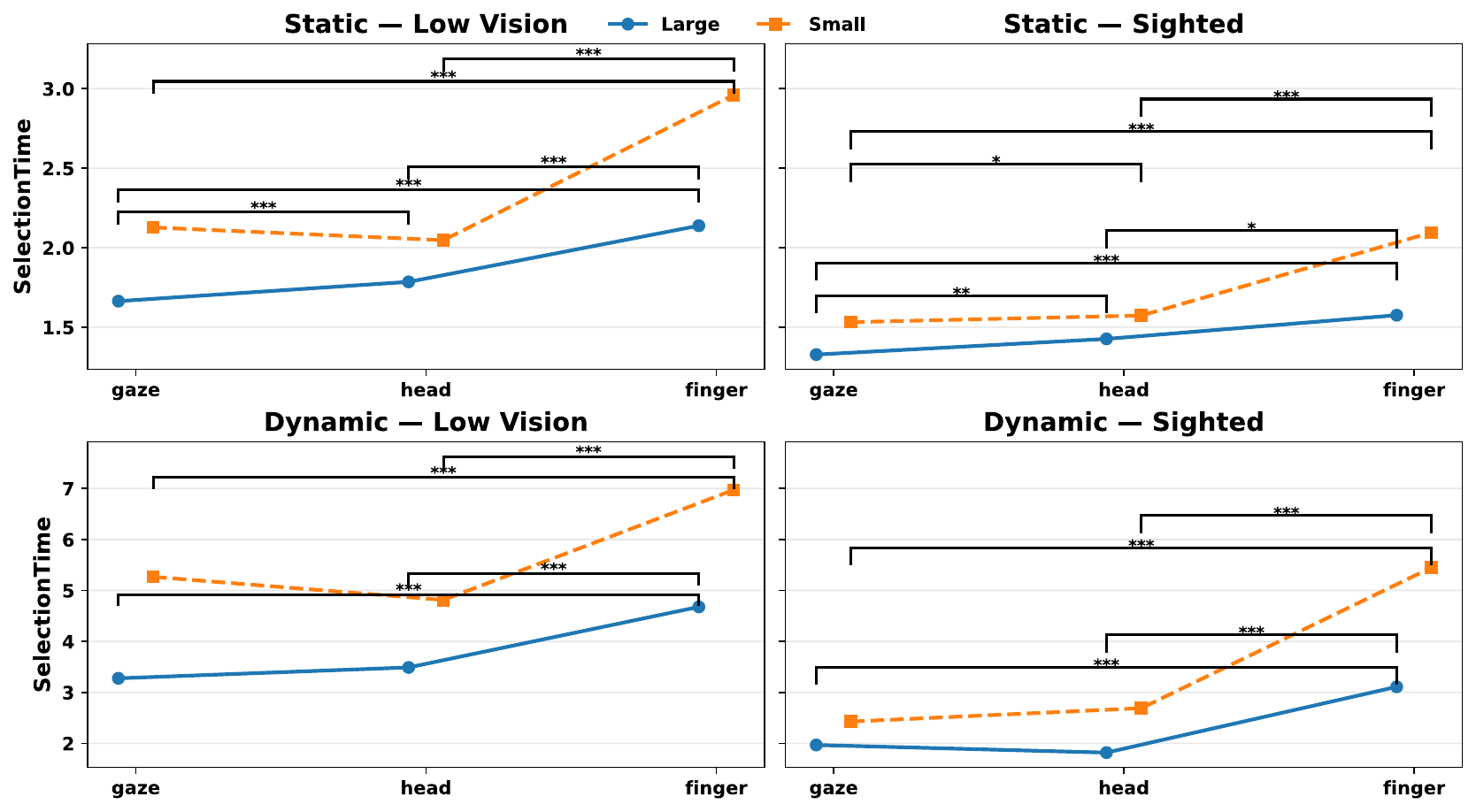}
    \end{subfigure}
    \begin{subfigure}{0.49\textwidth}
        \centering
    \includegraphics[width=\textwidth]{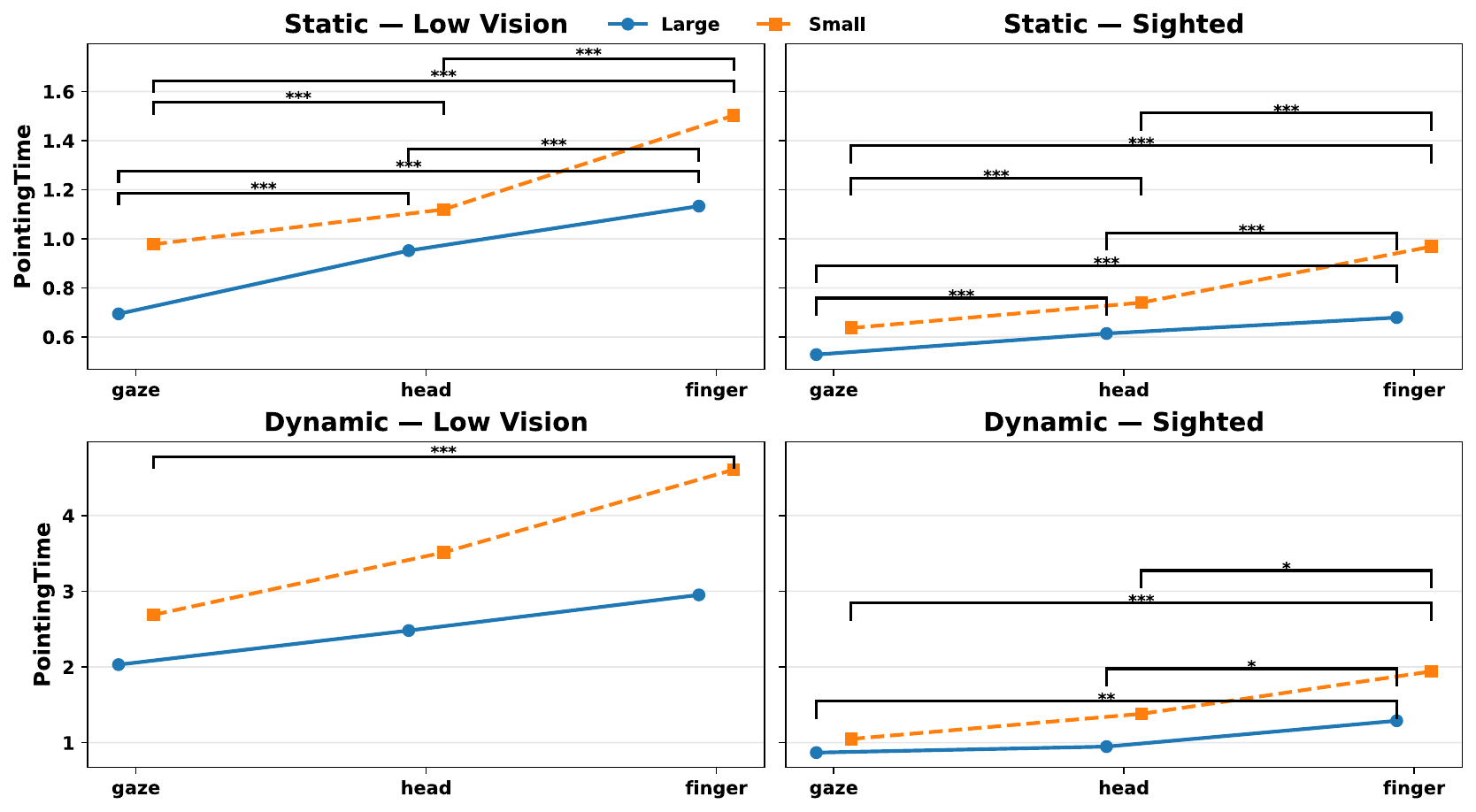}
    \end{subfigure}
    \caption{Mean \textit{SelectionTime} and \textit{PointingTime} across different techniques, target sizes, and scenarios. Orange color represents performance on small targets, while blue represents large targets. In the stationary scenario, gaze-based selection consistently achieved the fastest initial target landing and overall selection time for sighted users and for PVL selecting large targets. However, this advantage diminished under walking conditions, and for PLV selecting small targets in the stationary scenario, with gaze and head showing comparable performance. *** $p < .001$, ** $p < .01$, * $p < .05$. }
    \label{fig:efficiency}
    \vspace{-2ex}
\end{figure*}

\subsubsection{Selection Efficiency.} \label{sec:efficiency} We first report participants' selection efficiency through the overall selection time and pointing time in both stationary and dynamic scenarios.  

\textbf{\textit{Stationary Scenario.}} We found a significant main effect of \textit{Vision} ($F_{1,34} = 30.5/28.3$, $p <.001$, $\eta _p^2 =.47/.45$) on both \textit{SelectionTime} and \textit{PointingTime}, with low-vision users spending significantly longer time than sighted participants ($est.=362/368$, $p<.001$). For \textit{PointingTime}, we also found a significant \textit{Technique} main effect ($F_{2,1682} = 206.1$, $p < .001$, $\eta _p^2 =.20$), as shown in Figure \ref{fig:efficiency} top-right. Post-hoc comparisons showed that gaze-based selection enabled the fastest initial target landing ($est._{finger-gaze}=469$, $p<.001$; $est._{gaze-head}=-241$, $p<.001$), 
while finger-based was consistently the slowest ($est._{finger-head}=228$, $p<.001$).

When examining \textit{SelectionTime}, we found a significant three-way interaction effect ($F_{2,1682} = 6.2$, $p =.002$, $\eta _p^2 =.02$), indicating different technique performance across \textit{Vision} and \textit{Size} (see Figure \ref{fig:efficiency} top-left). For sighted users as well as PLV who selecting large targets, we saw that gaze achieved the best performance, followed by head, and finger selection behaved the slowest (all $p \leq 0.034$). However, when PLV selected small targets, no significant differences were found between gaze- and head-based selection ($est._{gaze-head}=-139$, $p=.085$), 
while finger-based selection remained the slowest ($est._{finger-head}=317$, $p<.001$; $est._{finger-gaze}=456$, $p<.001$).

\textbf{\textit{Dynamic Scenario.}} 
We found a significant main effect of \textit{Technique} ($F_{2,386} = 108.5$, $p < .001$, $\eta _p^2 =.36$) on \textit{SelectionTime} during walking (see Figure \ref{fig:efficiency} bottom-left). Finger-based selection resulted in significantly longer selection time than both gaze- and head-based selection ($est._{finger-head}=128$, $p<.001$; $est._{finger-gaze}=120$, $p<.001$), while no significant difference was observed between gaze- and head-based selection ($est._{gaze-head}=8$, $p=.692$). A significant three-way interaction effect ($F_{2,386} = 3.3$, $p = .036$, $\eta_p^2 = .02$) revealed different effects of \textit{Size} across techniques and groups. For sighted participants, \textit{Size} did not affect gaze-based selection performance ($est. = -44$, $p = .254$), but the performance of head- and finger-based selection significantly dropped as the target size decreased (head: $est. = -101$, $p < .001$; finger: $est. = -111$, $p < .001$). In contrast, for PLV, size effects were significant across all techniques (gaze $est. = -81$, $p < .001$; head $est. = -59$, $p = .020$; finger $est. = -78$, $p < .001$). Further, \textit{Vision} had a significant effect in several conditions, with PLV spending more time than sighted participants in gaze-based selection (both large and small targets) and head-based selection with large targets (all $p\leq.022$).

For \textit{PointingTime}, we also found a significant \textit{Technique} $\times$ \textit{Vision} $\times$ \textit{Size} interaction effect ($F_{2,386} = 3.0$, $p = .049$, $\eta_p^2 = .02$), as shown in Figure \ref{fig:efficiency} bottom-right. For sighted participants, finger-based selection resulted in significantly longer initial target landing time than both gaze- and head-based selection (all $p\leq.015$). For PLV, no significant differences between techniques were found with large targets (all $p\geq.261$), whereas with small targets, finger-based selection resulted in significantly longer time than gaze-based selection ($est. = 94$, $p <.001$). The effect of \textit{Size} was significant for sighted participants with head-based selection, and for both sighted and PLV with finger-based selection (all $p\leq.005$). Additionally, PLV required more time than sighted participants at the pointing stage for gaze- and head-based selection across target sizes (all $p\leq.044$), but not for finger-based selection, possibly reflecting a floor effect \cite{liu2021t,wang2008investigating}, where performance was already constrained by balance and arm movement coordination in walking ~\cite{pozzo2002coordination}.

\textbf{\textit{Interpretations:}} 
In the stationary scenario, gaze-based selection consistently achieved the fastest initial target landing and overall selection time for sighted users and for PVL selecting large targets. However, this advantage diminished under walking conditions, and for PLV with small targets in the stationary scenario, with gaze and head showing comparable performance. 

Additionally, in the dynamic scenario, for \textit{SelectionTime}, gaze-based selection was less sensitive to target size for sighted participants, whereas this advantage did not persist for PLV. For \textit{PointingTime}, both gaze- and head-based selection were less sensitive to target size for PLV, while only gaze-based selection showed reduced size sensitivity for sighted participants.

These findings highlighted the potential of gaze-based selection for both sighted and low-vision users. However, reduced target size and increased movement variability imposed greater stability demands in gaze control, especially for low-vision users, reducing the performance gap between gaze- and head-based selection and making head-based selection to be another feasible option for PLV in many conditions (e.g., small targets in stationary scenario, and dynamic scenario).

\subsubsection{Selection Stability.}
\label{subsubsec:sighted_PLV_stability}
We report selection stability through confirmation time and re-entry times during dwelling to reflect PLV's ability in dwelling on targets with different input modalities. 

\textbf{\textit{Stationary Scenario.}} We found no significant effects of \textit{Vision} on both \textit{ConfirmationTime}  and \textit{ReentryTimes}. For both measures, we found significant interaction effects of \textit{Technique} $\times$ \textit{Size} ($F_{2,1682} = 88.8/23.6$, $p < .001$, $\eta _p^2 =.10/.03$), indicating that target size affected technique performance differently. With gaze-based and finger-based dwelling, small targets took longer for users to confirm a selection and also resulted in more re-entries than large targets (gaze: $est. = -129/-194$, $p <.001$; finger: $est. = -224/-221$, $p <.001$), but no significant difference was observed between target sizes for head-based confirmation. 

\begin{figure*}[]
    \centering
    \begin{subfigure}{0.49\textwidth}
        \centering
    \includegraphics[width=\textwidth]{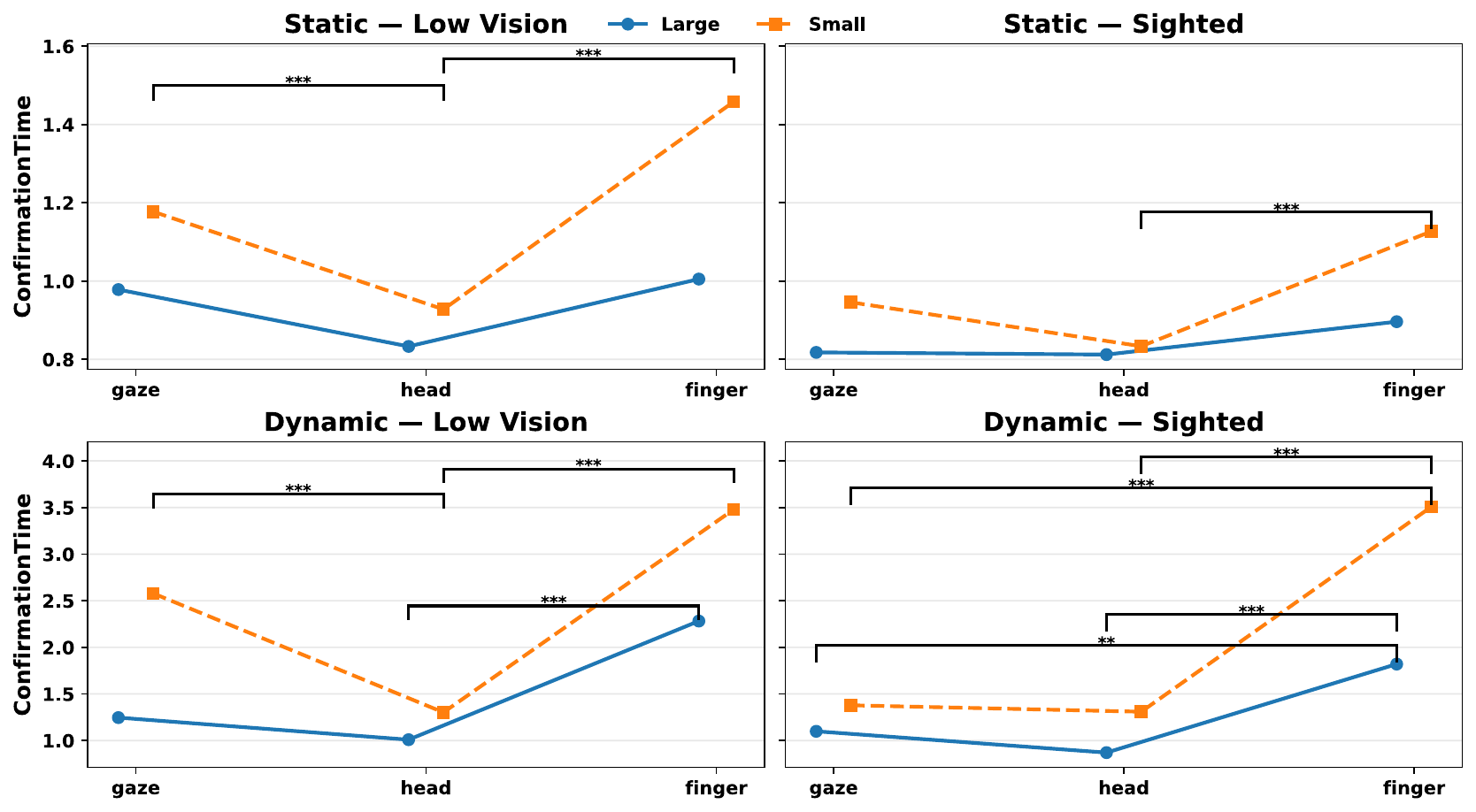}
    \end{subfigure}
    \begin{subfigure}{0.49\textwidth}
        \centering
    \includegraphics[width=\textwidth]{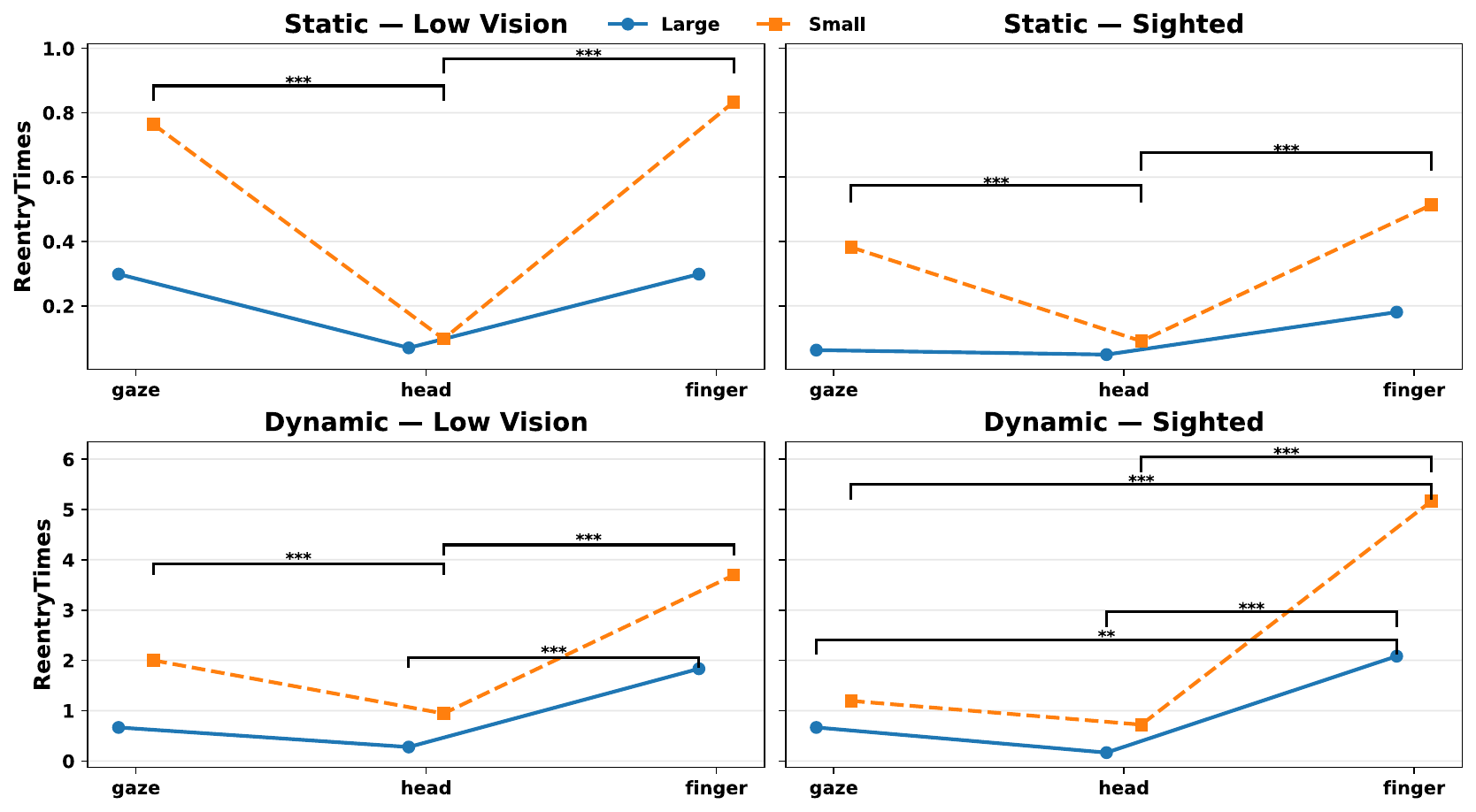}
    \end{subfigure}
    \caption{Mean \textit{ConfirmationTime} and \textit{ReentryTimes} across techniques, target sizes, and scenarios. Head-based dwelling remained the most stable across all conditions, whereas finger-based dwelling the least. The advantage of head-based dwelling over gaze was especially pronounced for PLV when selecting small targets. *** denotes $p < .001$, and ** denotes $p < .01$.}
    \label{fig:stability}
\end{figure*}

We also found significant three-way interactions of \textit{Technique} $\times$ \textit{Size} $\times$ \textit{Vision} on both measures ($F_{2,1682} = 88.8/74.4$, $p < .001$, $\eta _p^2 =.10/.08$), as shown in Figure \ref{fig:stability} top row. For small targets, head-based confirmation resulted in significantly shorter time than finger-based confirmation for sighted participants ($est. = 179$, $p <.001$), and shorter than both finger- and gaze-based confirmation for PLV ($est._{finger-head} = 267$, $p <.001$; $est._{gaze-head} = 152$, $p =.004$). Head-based also showed significantly fewer re-entries than both finger- and gaze-based confirmation for both groups (sighted/PLV: $est._{finger-head} = 204/300$, $p <.001$; $est._{gaze-head} = 129/258$, $p =.017/<.001$). However, for large targets, no significant differences between technique performance were observed across groups for both measures (all $p\geq.097$).

\textbf{\textit{Dynamic Scenario.}} We found no main effect of \textit{Vision} on either \textit{ConfirmationTime} or \textit{ReentryTimes}. 
However, a significant three-way interaction effect on \textit{ConfirmationTime} ($F_{2,386} = 10.7$, $p <.001$, $\eta _p^2 =.05$) indicated different effects of \textit{Vision} on performance across techniques and target sizes. Specifically, PLV required more time than sighted participants to confirm a selection on small targets when using gaze ($est.=90$, $p=.014$), but there is no difference between sighted and low-vision users when dwelling with finger or head, or dwelling on large targets (all $p\geq$.982).

The three-way interaction also indicated different effects of \textit{Size} across techniques and user groups. For \textit{ConfirmationTime}, small targets significantly increased confirmation time for PLV with gaze ($est.=110$, $p<.001$), but there is no such effect for other technique-vision combinations (all $p\geq.052$). For \textit{ReentryTimes}, a significant three-way interaction ($F_{2,386} = 5.3$, $p = .005$, $\eta _p^2 =.03$) showed size effects  for PLV using gaze- or finger-based confirmation (gaze: $est.=95$, $p=.002$; finger: $est.=81$, $p=.019$), but not with head-based confirmation, nor for sighted participants (all $p\geq.066$). 

Technique performance also differed across groups and target sizes for both measures, as shown in Figure \ref{fig:stability} bottom row. For sighted participants, finger-based confirmation resulted in longer time and more re-entries than both gaze- and head-based confirmation (all $p\leq.013$). For PLV, with large targets, head-based confirmation resulted in shorter time ($est.=108$, $p<.001$) and fewer re-entries than finger-based confirmation ($est.=105$, $p<.001$), whereas with small targets, head-based confirmation outperformed both gaze- and finger-based confirmation (all $p<.001$).

\textbf{\textit{Interpretations:}} Overall, we found that PLV and sighted participants showed comparable selection stability in both tasks, except when dwelling on small targets with gaze in the dynamic task. Head-based dwelling remained the most stable across tasks, user groups, and target sizes, while finger-based technique showed the lowest stability, with gaze-based dwelling staying in-between. However, the stability advantage of head- over gaze-based dwelling was reduced for sighted participants and for large targets, but remained pronounced for PLV when selecting small targets.

\subsubsection{Impact of Walking on Selection.}\label{subsubsec:walking_impact}

For \textit{WalkingTime}, we found significant main effects of \textit{Technique} and \textit{Vision} (both $p<.001$, $\eta _p^2 =.22/.31$). Participants walked more slowly with finger-based selection than with head- or gaze-based selection ($est._{finger-gaze}=69$, $p<.0001$; $est._{finger-head}=-73$, $p<.001$; $est._{gaze-head}=4$, $p=.867$), and PLV walked more slowly than sighted participants ($est.=118$, $p<.001$).

For \textit{SelectionTimeInWalking}, \textit{Technique} showed a significant main effect ($F_{2,386} = 101.9$, $p < .001$, $\eta _p^2 =.35$), following a pattern consistent with that observed for \textit{WalkingTime}: finger-based selection yielded the slowest performance ($est._{finger-gaze}=120$, $p<.001$; $est._{finger-head}=126$, $p<.001$), and no significant difference was found between gaze- and head-based selection ($est._{gaze-head}=6$, $p=.833$). No main effect of \textit{Vision} was found ($p=.108$). A significant \textit{Technique} $\times$ \textit{Vision} $\times$ \textit{Size} interaction revealed different size sensitivities across techniques and groups ($F_{2,386} = 4.8$, $p = .009$, $\eta _p^2 =.02$): sighted participants showed no size effect when using gaze-based selection ($est.=-29$, $p=.915$), but significant size effects for head- and finger-based selection (head: $est.=-86$, $p<.001$; finger: $est.=-97$, $p<.001$). In contrast, PLV showed significant size effects with gaze- and finger-based selection (gaze: $est.=-88$, $p<.001$; finger: $est.=-79$, $p<.001$), and a marginal effect for head-based selection ($est.=-60$, $p=.050$).

\textbf{\textit{Interpretations:}} Although PLV walked more slowly overall, selection imposes comparable performance costs on walking for both PLV and sighted participants. Gaze- and head-based selection achieved comparable high performance for both user groups, while finger-based selection consistently led to lower performance, suggesting higher coordination demands during walking. For sighted participants, gaze-based selection remained stable in performance during walking across target sizes. In contrast, PLV showed more sensitivity to target size during walking across all three techniques.

\subsubsection{Subjective Experiences}
\label{subsubsec:User Experiences Across Selection Techniques}
We report participants' subjective experiences to explain their statistical performance. 
We saw a significant effect of \textit{Technique} on all subjective measures, including perceived \textit{Performance}, \textit{Effectiveness}, \textit{Mental Demand}, \textit{Physical Demand}, and \textit{Frustration} (all $p < .001$), with no effect of \textit{Vision} (all $p\geq.100$).
While gaze input generally enabled the fastest overall performance statistically (Section \ref{sec:efficiency}), we found that head-based selection was rated to be most successful, effective, least frustrating, followed by gaze-based selection, with finger-based selection being rated the worst (all $p\leq.033$). We elaborate on participants' experiences below.

\textbf{\textit{Stability.}} Head-based selection was frequently described as stable (12/20 PLV; 7/18 sighted) and accurate (6; 10)\footnote{Parenthetical values follow the format (PLV; sighted), with omitted denominators indicating counts out of 20 and 18, respectively.}. Finger-based selection was frequently described as unstable (13; 16), particularly over longer distances (3; 1). While instability in gaze dwelling was reported by both groups (13; 12), participants attributed their performance to different reasons: more sighted participants reported gaze cursor inaccuracies (6; 14) and attributed them to limitations of the eye-tracking technology (2; 6), whereas more PLV attributed them to their own visual abilities (9; 1). As L05 reflected, ``I couldn't seem to get [the cursor] to stay on the target ... I don't know if that's a function of how my eyes work. Do I have nervous eyes?''

\textbf{\textit{Ease of Control.}} Consistently, participants described head-based selection as easy to control (17/20 PLV; 16/18 sighted) with minimal physical effort (13; 12). However, some participants noted increased effort from headset weight during head-based selection (2; 2), particularly for higher targets (L02). Perceptions of gaze were mixed (difficult: 8; 7, easy: 7; 9), though most participants (15; 16) reported low physical effort ($rating\leq3$). A few participants noted eye fatigue during sustained gaze use (1; 4). Finger-based selection was often considered difficult and physically demanding (13; 13).

\textbf{\textit{Flexibility.}}
Participants reported that gaze-based selection required sustained and focused attention (9/20 PLV; 12/18 sighted), with four PLV
noting that this increased attentional demand could raise safety concerns during walking. Dwell-based confirmation may therefore be unsuitable for gaze interaction in mobile contexts (see Section~\ref{subsec:design_considerations}). In contrast, head-based selection was perceived as requiring less focused attention (6; 6), with finger-based selection generally falling in between. Several participants (L16, L18, S01) further commented that gaze-based selection resulted in a more limited functional field of view compared to head-based selection, constraining flexibility.

\textbf{\textit{Naturalness.}}
Gaze- and head-based selection were often described as natural because the cursor aligned with users’ attention (gaze: 11/20 PLV, 13/18 sighted; head: 10, 13). Head-based selection was particularly natural to participants with peripheral vision loss (L02, L17, L18) as it aligned with their original viewing strategy. As L18 noted, he sometimes needed to adjust his head when using gaze-based technique to prevent targets from falling into blind spots, making head-based selection feel more natural. 
Finger-based selection received mixed evaluations, with similar counts between natural (1; 5) and unnatural (4; 4). L08 and L10 expressed a preference for directly touching targets rather than pointing with mid-air gestures. Several participants reported that head- and finger-based techniques required an additional refinement step to align the cursor, compared to gaze-based selection (head: 5, 6; finger: 9, 10).

For real-world usability, gaze- and head-based techniques were appreciated for their discretion (L13, L20, S16), whereas finger-based selection was sometimes perceived as socially awkward (L11, L13) or impractical, particularly for PLV using white canes (L06).

\textbf{\textit{Cursor challenges for PLV.}} While no sighted participants reported cursor-specific challenges, seven PLV reported that the cursor was difficult to track during the dynamic scenario, particularly over long distances (6 PLV) and in visually complex environments (L15). Participants suggested increasing cursor size (L02, L04, L09), brightness (L19), and contrast (L15). However, some (L05, L06) were concerned that overly large cursors might occlude small targets, indicating a trade-off between visibility and precision.

Cursor challenges varied across techniques. Four PLV (L06, L09, L16, L17) found the gaze cursor easy to track.
L06 found that the gaze cursor helped him focus: ``It's kind of like having a second arm of your eye. When you have that [cursor] there, that's helping you focus on [the target].'' However, L10, who experienced double vision, reported seeing two gaze cursors when not fully focused, which he found distracting. Similarly, five low-vision participants felt the gaze cursor was difficult to track due to stability issues.

For head-based selection, four participants (L03, L13, L14, L20) reported that the head cursor was easy to locate, and L10 noted that head movement did not trigger double vision. However, L19 mentioned that the head cursor could still fall into blind spots when the eyes shifted, sharing: ``I'd lose the target for a second. But I knew exactly where [the cursor] was because it was responding to my head movement only.''

Finger-based cursors were most frequently described as difficult to track (10/20). L15 and L17 observed that while the head cursor generally remained within the visible region of their visual field, the finger cursor often moved into blind spots. Additionally, due to inherent hand instability during mid-air pointing, L06 and L08 described the cursor movement as ``distracting.'' As L06 noted, ``When your finger even like twitches just a little bit, it kind of throws that off from where you're trying to look.''

\subsection{Impact of Visual Conditions on Selection}
In this section, we report the effects of different low-vision conditions (i.e., visual acuity, peripheral vision, central vision) on PLV’s selection performance and experiences.

\subsubsection{Effects of Low-Vision Conditions on Selection Efficiency}
\label{subsubsec:LA and PVL} We examined the effects of \textit{VisualAcuity} and \textit{PeripheralVision} on selection efficiency. In the stationary scenario, we found significant effects of \textit{Technique} $\times$ \textit{VisualAcuity} $\times$ \textit{PeripheralVision} on \textit{SelectionTime} ($\chi^2(2)=9.0$, $p=.011$) and \textit{PointingTime} ($\chi^2(2)=15.9$, $p<.001$). \textit{LowAcuity} significantly increased head- and gaze-based \textit{SelectionTime} (head: $p = .002$; gaze: $p = .020$) as well as head-based \textit{PointingTime} ($p=.008$) among participants with \textit{SevereFieldLoss}, but no difference was found in the \textit{MildFieldLoss} group (all $p\geq.537$). Similarly, \textit{SevereFieldLoss} increased the \textit{SelectionTime} of all techniques (head: $p=.003$; gaze: $p=.017$; finger: $p =.043$) as well as the \textit{PointingTime} of head- and finger-based selection (head: $p=.009$; finger: $p=.028$) among \textit{LowAcuity} participants, but no difference was found among \textit{HighAcuity} participants (all $p\geq.319$).

In the dynamic scenario, we found significant effects of \textit{VisualAcuity} $\times$ \textit{PeripheralVision} on \textit{SelectionTime} and \textit{PointingTime} ($\chi^2(1)=11.7/11.3$, $p<.001$), with participants with both \textit{LowAcuity} and \textit{SevereFieldLoss} exhibiting the longest pointing and overall selection time (all $p \leq .015$).

\textbf{\textit{Interpretation.}} These results suggest that reduced visual acuity and peripheral vision loss together produced the greatest selection difficulty. Notably, in the stationary scenario, gaze still showed an advantage at the pointing stage by remaining comparable in performance across low-vision conditions, whereas both head- and finger-based selections were more affected by severe low vision.

\subsubsection{Effects of Low-Vision Conditions on Selection Stability} We examined the effects of \textit{VisualAcuity} and \textit{PeripheralVision} on selection stability. In the stationary scenario, we found a significant three-way interaction effect of \textit{Technique} $\times$ \textit{VisualAcuity} $\times$ \textit{PeripheralVision} on \textit{ConfirmationTime} ($\chi^2(2)=6.8$, $p=.034$). \textit{LowAcuity} significantly increased confirmation time for participants with \textit{SevereFieldLoss} when using gaze- and head-based dwelling (head: $p=.034$; gaze: $p=.004$), but no difference was seen in the \textit{MildFieldLoss} group (all $p\geq.506$). \textit{SevereFieldLoss} significantly increased \textit{ConfirmationTime} among participants with \textit{LowAcuity} when using gaze- and head-based dwelling (head: $p=.028$; gaze: $p=.024$), but no difference was found among the \textit{HighAcuity} group (all $p\geq.405$).
For \textit{ReentryTimes}, we found a significant effect of \textit{VisualAcuity} $\times$ \textit{PeripheralVision} ($\chi^2(1)=5.0$, $p=.025$),
showing that re-entry times increased only when \textit{LowAcuity} and \textit{SevereFieldLoss} co-occurred (\textit{LowAcuity}: $p_{Mild-Severe}=.015$; \textit{SevereFieldLoss}: $p_{Low-High}=.012$).

In the dynamic scenario, we found significant main effects of both \textit{VisualAcuity} and \textit{PeripheralVision} on \textit{ConfirmationTime} ($\chi^2(1)=8.2/6.5$, $p=.004/.011$). \textit{LowAcuity} and \textit{SevereFieldLoss} both increased confirmation time ($p=.019/.035$).
For \textit{ReentryTimes}, no significant effect of \textit{PeripheralVision} was observed ($p=.167$). However, a significant effect of \textit{Technique} $\times$ \textit{VisualAcuity} ($\chi^2(2)=11.1$, $p=.004$) suggested \textit{LowAcuity} significantly increased \textit{ReentryTimes} for only head-based confirmation ($p<.001$).

\textbf{\textit{Interpretation.}} We again found compounded effects of visual constraints in selection stability. However, finger-based confirmation in the stationary scenario (\textit{ConfirmationTime}) and gaze- and finger-based confirmation in the dynamic scenario (\textit{ReentryTimes}) did not show increased instability under \textit{LowAcuity} or \textit{SevereFieldLoss}. This may be due to a ceiling effect \cite{liu2021t,wang2008investigating}, where these tasks were already difficult for PLV regardless of visual ability.

\subsubsection{Unique Preferences of Participants with Central Vision Loss (CVL)} \label{subsubsec:CVL}
Four participants experienced CVL (L13-L15, L20) and relied on their preferred retinal locus (PRL) to view information (Section \ref{subsec:gaze_cursor_adjustment}). They showed different interaction preferences from other PLV. Unlike most participants who preferred head- or gaze-based selection, L14 and L20 who had both central and peripheral vision loss rated finger-based selection as their preferred technique. L13, L14, and L20 also reported easier control with finger-based interaction. L14 found head-based selection difficult as it required an additional step with CVL: he first needed to locate the target using his PRL and then align the head cursor. In contrast, he perceived finger-based selection as a more direct, single-step interaction.

Regarding gaze-based selection, L14 and L15 shared that aligning the gaze cursor with their PRL (Section \ref{subsec:gaze_cursor_adjustment}) improved eye tracking usability. However, they still needed additional adjustment when controlling the cursor in the PRL (L13–L15) due to the use of multiple PRLs (e.g., when the PRL that contained the cursor was not the one that located the target). L15 noted that gaze use in PRL became more natural with practice: ``It just felt more comfortable [after some trials] and I just felt more natural to walk and to gaze.''

%% file: sections/5-discussion.tex
\section{Discussion}
In this paper, we explored how people with low vision (PLV) perform with and perceive head-, gaze-, and finger-based selection techniques, and how variations in visual acuity and peripheral vision shape their performance and experiences. We found that for PLV, gaze-based selection enabled the fastest initial pointing when sitting and comparable overall selection time to head-based selection in both scenarios. However, PLV experienced greater difficulty in gaze-based dwelling compared to sighted participants, whereas head-based dwelling remained the most robust across target sizes, scenarios, and visual abilities (\textit{RQ1}). We also observed a compounded effect of the combination of reduced acuity and restricted field of view on both selection efficiency and stability. Moreover, while finger-based techniques resulted in the lowest performance for all sighted and most low-vision participants, participants with central vision loss found it useful due to the difficulty of controlling gaze or head cursor with their PRL (\textit{RQ2}).

In this section, we discuss the potential of gaze-based selection for PLV, derive design implications for accessible object selection in AR, and reflect on real-world challenges of AR selection.

\subsection{Gaze-Based Selection for Low Vision}
\label{subsec:gaze-based for PLV}
One important contribution of our research is to demonstrate the feasibility and advantage of gaze-based selection for PLV, despite their potential vision and gaze control challenges. 
These findings highlight the potential of incorporating eye tracking into future selection-based augmentation systems for PLV, offering benefits in pointing efficiency, intuitiveness, and discretion. While prior work has extensively investigated gaze-based techniques for sighted users \cite{kim2025pinchcatcher,chen2023gazeraycursor,wagner2024gaze,shi2023exploring}, our work was the first that focused on people with diverse low-vision conditions.

Despite its potential, our findings also reveal practical challenges for PLV in gaze-based selection. First, appropriate calibration support is a prerequisite. With high-contrast, large calibration targets and explicit guidance on target positions \cite{wang2023understanding, wang2024gazeprompt}, PLV were able to complete the eye tracking calibration. For users with central vision loss, moving the cursor to their PRL (instead of central vision) enables successful use of gaze-based selection. Although some participants reported initial difficulty controlling the cursor with their PRL, L15 described noticeable improvement with practice (Section~\ref{subsubsec:CVL}), consistent with prior work that a trained retinal locus can be developed for a specific task \cite{oflaz2022short, watson2006effects, nilsson1998location}. These findings suggest that future AR systems should provide customizable, guided calibration interfaces with cursor alignment options, to support accessible gaze-based interaction for PLV.

While gaze exhibited faster pointing, gaze-based dwelling showed reduced stability especially for PLV, suggesting a trade-off. One possible explanation is that gaze-based selection directly maps subtle eye movements to the cursor position, making small misalignments more noticeable during interaction. This issue could be more pronounced for people with low visual acuity as they spend more visual efforts to identify cursor position and target boundaries. Future systems should explore gaze smoothing techniques adaptive to low-vision conditions to improve cursor stability. In addition, clearly communicating tracking accuracy and calibration limits may help users better handle misalignment and reduce self-blame.

\subsection{Design Implications for Accessible Object Selection in AR} \label{subsec:design_considerations}
We discuss potential design implications for future AR systems to support accessible object selection.

\textbf{Providing multiple selection options.} While sighted participants and PLV generally shared preferences for head- and gaze-based selection, several participants with both peripheral and central vision loss favored finger-based selection. This variation suggests that no single technique fits all users with diverse visual conditions. Future AR systems should therefore support multiple selection techniques and allow customization.

\textbf{Head as the default option for PLV.} Across target sizes and task scenarios, head-based selection showed consistently stable performance for PLV. While gaze-based selection demonstrated advantages in certain conditions (e.g., large targets in stationary scenarios), it was less stable in dwelling, was subjectively less preferred among PLV, and required additional calibration. These findings suggest that head-based selection can serve as a robust default option for AR object selection, particularly for PLV.

\textbf{Reconsidering Dwell-Based Confirmation.} 
Dwell-based triggering (i.e., maintaining the cursor stably within a target) poses challenges for PLV in gaze-based interaction. Increased movement in dynamic scenarios (e.g., walking) further perpetuates the stability issue. Future research should explore alternative confirmation mechanisms to compensate for the weakness of gaze-based dwelling. Gaze gestures (e.g., directional gaze movements \cite{mohan2018dualgaze} or leave-and-return patterns \cite{hyrskykari2012gaze}) could be considered but the design should be mindful of PLV's visual abilities; for example, blinking gestures may not be suitable as some PLV tend to blink more due to eye dryness \cite{abbott2024prevalence,nakamori1997blinking,rahman2015corneal}. Other confirmation modalities could also be considered after gaze pointing, such as wristband gestures~\cite{metaWristbandsTechnology, truong2018capband}.

\textbf{Adaptive Cursor Design.}
Increased cursor visibility for PLV can support orientation, but the extruded cursor design may obstruct small objects or affect scene understanding, particularly in gaze- and head-based techniques where the cursor occupies the central field.
Future work should explore adaptive cursor design in complex environments, dynamically adjusting size and color based on visual background. Additionally, cursor design could also adapt to user intent to reduce obtrusiveness, which may be inferred from gaze behavior (e.g., searching vs.\ observing) \cite{wang2025characterizing}. For example, the cursor can remain invisible during pursuit tasks but become visually pronounced for PLV during discrete selection \cite{wei2025reevaluating}.

\subsection{Real-world Selection Challenges}
This work sets a foundation for future selection-based AR augmentations for PLV by investigating feasible selection techniques in both stationary and dynamic scenarios. While our initial study was conducted in an ideal controlled lab setting, selection in the real world could be much more complex and challenging.  
For example, the environment may contain multiple crowded or even overlapping selectable objects; there could also be various dynamic situations beyond walking, such as sitting on a bus, that further hinder selection. Future systems should explore how to design more accessible selection techniques to overcome these real-world challenges, for example, integrating object augmentation \cite{zhao2016cuesee} and distinction techniques \cite{lee2024cookar} to help distinguish close or overlapping objects before further confirmation, or predicting intended targets in a broader selected area \cite{kim2025pinchcatcher} based on user intent or context.
It is also important to investigate when and how systems should intervene under ambiguous selection to request explicit clarification without interrupting user flow, particularly in dual-task or fast moving scenarios. Prior work has examined how voice assistants can solicit explicit user feedback \cite{xiao2021let}, and future research should explore how such design guidelines can be adapted and expanded to support PLV in selection-based AR augmentations.

\subsection{Limitations and Future Directions}
Our research has limitations. First, the experimental design was relatively simple and conducted in controlled settings. In real world, object selection may involve visual clutter, low contrast, occlusion, and ambiguous boundaries, which may pose additional challenges for PLV and further affect selection performance and stability. Our future work will design more complex studies to dive deep into these real-world challenges. Second, our work was not a traditional Fitts’ Law study. Instead of rendering standard virtual targets in tightly controlled directions and distances, we focused on selecting real-world objects in AR to reflect more realistic use scenarios. While this design better reflects practical AR use and experiences, it limits direct comparison with classic pointing models and does not allow us to establish a predictive performance model that a Fitts’ Law study typically does. Future work should examine how to refine the study design to develop predictive models for AR object selection while ensuring a reasonable task load for PLV.

%% file: sections/6-conclusion.tex
\section{Conclusion}
We contributed the first study investigating how PLV use head-, gaze-, and finger-based selection techniques for target selection in AR environments. We analyzed selection performance and user experiences between sighted participants and PLV, and across diverse low-vision conditions. Our results show that PLV achieve the fastest initial pointing with gaze but experience difficulty maintaining gaze-based dwelling stability, whereas head-based selection remains the most robust across visual conditions. We also identify a compounded effect of reduced acuity and restricted field of view on performance. Together, these findings provide empirical insights into suitable AR selection techniques for people with different visual conditions, inspiring future selection-based visual augmentations.

%% file: sections/Appendix.tex
\clearpage
\onecolumn

\section{Demographic Information of Low Vision Participants} \label{sec:demographics_plv}

\begin{table*}[ht]
\centering
\tiny
\begin{tabular}{>{\centering\arraybackslash}m{0.3cm} >{\centering\arraybackslash}m{0.6cm} >{\centering\arraybackslash}m{1.8cm} > {\centering\arraybackslash}m{0.5cm} >{\centering\arraybackslash}m{1.8cm} >{\centering\arraybackslash}m{1cm} >{\centering\arraybackslash}m{4cm} > {\centering\arraybackslash}m{2cm} >
{\centering\arraybackslash}m{0.8cm} > {\centering\arraybackslash}m{0.5cm}}
\Xhline{2\arrayrulewidth}
\textbf{ID} & \textbf{Age/ Gender} & \textbf{Diagnosis} & \textbf{Legally Blind} & \textbf{Visual Acuity (After Correction)}& \textbf{Low v.s. High Acuity} & \textbf{Field of View (FOV)} & \textbf{Severe v.s. Mild Peripheral Vision Loss} & \textbf{Prior AR Experiences} & \textbf{Dominant Hand} \\
\Xhline{2\arrayrulewidth}
L01 & 59/M & Optical neuropathy & Y & R: 20/300; L:20/70 & High Acuity &Less than 20°&Severe Peripheral Field Loss &  Y &Right\\
\hline
L02 & 23/Non-binary & Unsure & N & R: 20/80; L:20/80 & High Acuity& $\sim$60°&Severe Peripheral Field Loss& Y & Right\\
\hline
L03 & 62/M & Birth defect & N & R: <20/400; L:20/50 & High Acuity& Intact &Mild Peripheral Field Loss & Y & Right \\
\hline
L04 & 57/F & Macular degeneration & N & R: 20/320; L:20/125 & Low Acuity&Intact &Mild Peripheral Field Loss& Y & Left \\
\hline
L05 & 67/M & Sleep apnea with suspected optic nerve impact& N& R: 20/60; L:20/15 &High Acuity& Reported two-thirds of the lower vision loss in right eye; passed all visual field test points&Mild Peripheral Field Loss & Y &Right \\
\hline
L06 & 41/M & Central scotoma & Y & R: 20/200; L: 20/200 &Low Acuity& Intact &Mild Peripheral Field Loss & Y & Right \\
\hline
L07 & 59/F & Central scotoma in the right eye
 & N & R: 20/40; L: 20/30 &High Acuity& Reported scotoma; passed all visual field test points &Mild Peripheral Field Loss & N & Right\\
\hline
L08 & 19/M & Stargardts & N & R: 20/160; L: 20/250 &Low Acuity& Intact &Mild Peripheral Field Loss & N & Right\\
\hline
L09 & 58/F & Retinitis pigmentosa & Y & R: 20/80; L: 20/60 &High Acuity&$\sim$10°& Severe Peripheral Field Loss & N & Right\\
\hline
L10 & 29/M & Stargardts & N & R: 20/100; L: 20/100 &Low Acuity& Intact &Mild Peripheral Field Loss& Y & Right \\
\hline
L11 & 72/F & Macular degeneration; chemotherapy-induced neurological corneal issues & N & R: 20/150; L: 20/200 &Low Acuity& Intact &Mild Peripheral Field Loss & Y & Right\\
\hline
L12 & 78/F & Glaucoma in both eyes; ICE syndrome in right eye & N & R: 20/160; L: 20/30 &High Acuity & Reported peripheral vision loss in the right eye ($\sim$90°); passed all visual field test points&Mild Peripheral Field Loss & N & Right\\
\hline
L13 & 50/M & Macular degeneration (Best disease) & Y & R: 20/600--20/800; L: 20/400 &Low Acuity& Central scotomas &Mild Peripheral Field Loss & Y & Right\\
\hline
L14 & 81/M & Corneal dystrophy & Y & R: <20/400; L:< 20/400 &Low Acuity& Central scotomas and peripheral field loss in the upper and left visual fields (less than 50° remaining)&Severe Peripheral Field Loss& N & Right\\
\hline
L15 & 74/M & Stargardts & Y & R: <20/400; L: <20/400 &Low Acuity& Central scotomas&Mild Peripheral Field Loss & Y & Right \\
\hline
L16 & 42/M & Retinitis pigmentosa & Y & R: 20/40; L: 20/30 &High Acuity&$\sim$15°& Severe Peripheral Field Loss & Y & Right \\
\hline
L17 & 64/F & Glaucoma; cataract & N & R: 20/60; L: 20/60 &High Acuity&$\sim$40°& Severe Peripheral Field Loss & N & Left\\
\hline
L18 & 77/M & Glaucoma in the right eye; fluid behind the retina in the left eye, resulting in blindness & N & R: 20/125; L: <20/400 &Low Acuity&Left eye blindness with scotomas in the right eye& Severe Peripheral Field Loss
 & N & Right\\
\hline
L19 & 91/F & Glaucoma; macular degeneration & N & R: <20/400, L: 20/200 &Low Acuity& Intact &Mild Peripheral Field Loss & N & Right\\
\hline
L20 & 70/F & Stargardts & Y & R: 20/600; L: 20/600 &Low Acuity& Central scotomas, peripheral vision loss in the upper visual field, peripheral scotomas& Severe Peripheral Field Loss & Y & Right\\
\Xhline{2\arrayrulewidth}
\end{tabular}
\caption{Demographic information of participants with low vision. Low and high acuity were classified based on a 20/100 visual acuity threshold in the better eye. Severe Peripheral Field Loss here denotes peripheral vision loss with a visual field narrower than 60°, and Mild Peripheral Field Loss denotes mild to no peripheral vision loss with a visual field wider than 60°. Legally blind refers to best-corrected visual acuity worse than 20/200 in the better eye or a visual field narrower than 20°.}
\label{tab:demographics_plv}
\end{table*}

\section{Demographic Information of Sighted Participants}
\label{sec:demographics_sighted}
\begin{table}[H]
\centering
\scriptsize
\begin{tabular}{>{\centering\arraybackslash}m{0.5cm} >{\centering\arraybackslash}m{1.5cm} >{\centering\arraybackslash}m{2.5cm} > {\centering\arraybackslash}m{2cm}}
\Xhline{2\arrayrulewidth}
\textbf{ID} & \textbf{Age/ Gender} & \textbf{Prior AR Experiences} & \textbf{Dominant Hand} \\
\Xhline{2\arrayrulewidth}
S01 & 20/F  &  Y &Right\\
\hline
S02 & 27/F  &  Y &Right\\
\hline
S03 & 59/F  &  N &Right\\
\hline
S04 & 49/M &   N &Right\\
\hline
S05 & 23/M &   Y &Right\\
\hline
S06 & 24/F &  N &Right\\
\hline
S07 & 27/F & N &Right\\
\hline
S08 & 26/F &  N &Right\\
\hline
S09 & 44/M & Y &Right\\
\hline
S10 & 22/F & Y &Right\\
\hline
S11 & 57/F &  N &Right\\
\hline
S12 & 45/F & Y &Right\\
\hline
S13 & 21/M & Y &Right\\
\hline
S14 & 73/F & N &Left\\
\hline
S15 & 26/F & Y &Left\\
\hline
S16 & 41/M & Y &Left\\
\hline
S17 & 26/M & N &Right\\
\hline
S18 & 40/M & Y &Left\\

\Xhline{2\arrayrulewidth}
\end{tabular}
\caption{Demographic information of sighted participants.}
\label{tab:demographics_sighted}
\end{table}

\clearpage
\section{Statistical Results for Comparing Selection Behavior Between PLV and Sighted Participants} \label{sec:stats_plv_sighted}
\begin{table}[H]
\centering
\scriptsize
\begin{tabular}{>{\centering\arraybackslash}m{1cm} >{\centering\arraybackslash}m{1.8cm} >{\centering\arraybackslash}m{3cm} > {\centering\arraybackslash}m{2.5cm} > {\centering\arraybackslash}m{1cm} > {\centering\arraybackslash}m{1cm}}
\Xhline{2\arrayrulewidth}
\textbf{Task} & \textbf{Measure} & \textbf{Parameter} & \textbf{F} & \textbf{P-value} & \textbf{$\eta _p^2 $} \\
\Xhline{2\arrayrulewidth}
\multirow{28}{*}{Static} & \multirow{7}{*}{\textit{SelectionTime}} & \textit{Technique} &  $F(2,1682)=233.2$ & ***&0.22\dag \\
\cline{3-6}
 &  & \textit{Vision} &  $F(1,34)=30.5$ & ***&0.47\dag \\
\cline{3-6}
 &  & \textit{Size} &  $F(1,1682)=284.8$ & ***&0.14\dag \\
\cline{3-6}
 &  & \textit{Technique} x \textit{Vision} &  $F(2,1682)=40.2$ & ***&0.05 \\
\cline{3-6}
 &  & \textit{Technique} x \textit{Size} &  $F(2,1682)=48.1$ & ***&0.05 \\
\cline{3-6}
 &  & \textit{Vision} x \textit{Size} &  $F(1,1682)=29.7$ & ***&0.02\\
\cline{3-6}
 &  & \textit{Technique} x \textit{Vision} x \textit{Size} &  $F(2,1682)=6.2$ & **&<0.01 \\
\cline{2-6}
& \multirow{7}{*}{\textit{PointingTime}} & \textit{Technique} &  $F(2,1682)=206.1$ & ***&0.20\dag \\
\cline{3-6}
 &  & \textit{Vision} &  $F(1,34)=28.3$ & ***&0.45\dag \\
\cline{3-6}
 &  & \textit{Size} &  $F(1,1682)=149.7$ & ***&0.08\dag \\
\cline{3-6}
 &  & \textit{Technique} x \textit{Vision} &  $F(2,1682)=26.9$ & ***&0.03 \\
\cline{3-6}
 &  & \textit{Technique} x \textit{Size} &  $F(2,1682)=15.7$ & ***&0.02 \\
\cline{3-6}
 &  & \textit{Vision} x \textit{Size} &  $F(1,1682)=4.3$ & *&<0.01\\
\cline{3-6}
 &  & \textit{Technique} x \textit{Vision} x \textit{Size} &  $F(2,1682)=0.2$ & 0.809&-- \\
\cline{2-6}
& \multirow{7}{*}{\textit{ConfirmationTime}} & \textit{Technique} &  $F(2,1682)=85.9$ & ***&0.09\dag \\
\cline{3-6}
 &  & \textit{Vision} &  $F(1,34)=24.6$ & ***&0.42\dag \\
\cline{3-6}
 &  & \textit{Size} &  $F(1,1682)=214.1$ & ***&0.11\dag \\
\cline{3-6}
 &  & \textit{Technique} x \textit{Vision} &  $F(2,1682)=58.8$ & ***&0.07\dag \\
\cline{3-6}
 &  & \textit{Technique} x \textit{Size} &  $F(2,1682)=88.8$ & ***&0.10\dag \\
\cline{3-6}
 &  & \textit{Vision} x \textit{Size} &  $F(1,1682)=53.1$ & ***&0.03\\
\cline{3-6}
 &  & \textit{Technique} x \textit{Vision} x \textit{Size} &  $F(2,1682)=88.8$ & ***&0.10\dag \\
\cline{2-6}
& \multirow{7}{*}{\textit{ReentryTimes}} & \textit{Technique} &  $F(2,1682)=113.7$ & ***&0.12\dag \\
\cline{3-6}
 &  & \textit{Vision} &  $F(1,34)=15.9$ & ***&0.32\dag \\
\cline{3-6}
 &  & \textit{Size} &  $F(1,1682)=65.6$ & ***&0.04\\
\cline{3-6}
 &  & \textit{Technique} x \textit{Vision} &  $F(2,1682)=20.0$ & ***&0.02 \\
\cline{3-6}
 &  & \textit{Technique} x \textit{Size} &  $F(2,1682)=23.6$ & ***&0.03 \\
\cline{3-6}
 &  & \textit{Vision} x \textit{Size} &  $F(1,1682)=124.1$ & ***&0.07\dag\\
\cline{3-6}
 &  & \textit{Technique} x \textit{Vision} x \textit{Size} &  $F(2,1682)=74.4$ & ***&0.08\dag \\
\hline

\multirow{42}{*}{Dynamic} & \multirow{7}{*}{\textit{SelectionTime}} & \textit{Technique} &  $F(2,386)=108.5$ & ***&0.36\dag \\
\cline{3-6}
 &  & \textit{Vision} &  $F(1,34)=15.5$ & ***&0.31\dag \\
\cline{3-6}
 &  & \textit{Size} &  $F(1,386)=178.0$ & ***&0.32\dag \\
\cline{3-6}
 &  & \textit{Technique} x \textit{Vision} &  $F(2,386)=1.2$ & 0.316&--\\
\cline{3-6}
 &  & \textit{Technique} x \textit{Size} &  $F(2,386)=19.0$ & ***&0.09\dag \\
\cline{3-6}
 &  & \textit{Vision} x \textit{Size} &  $F(1,386)=10.9$ & **&0.03\\
\cline{3-6}
 &  & \textit{Technique} x \textit{Vision} x \textit{Size} &  $F(2,386)=3.3$ & *&0.02\\
\cline{2-6}
& \multirow{7}{*}{\textit{PointingTime}} & \textit{Technique} &  $F(2,386)=60.0$ & ***&0.24\dag \\
\cline{3-6}
 &  & \textit{Vision} &  $F(1,34)=28.7$ & ***&0.46\dag \\
\cline{3-6}
 &  & \textit{Size} &  $F(1,386)=106.9$ & ***&0.22\dag \\
\cline{3-6}
 &  & \textit{Technique} x \textit{Vision} &  $F(2,386)=8.5$ & ***&0.04 \\
\cline{3-6}
 &  & \textit{Technique} x \textit{Size} &  $F(2,386)=11.9$ & ***&0.06\dag \\
\cline{3-6}
 &  & \textit{Vision} x \textit{Size} &  $F(1,386)=30.1$ & ***&0.07\dag \\
\cline{3-6}
 &  & \textit{Technique} x \textit{Vision} x \textit{Size} &  $F(2,386)=3.0$ & *&0.02\\
\cline{2-6}
& \multirow{7}{*}{\textit{ConfirmationTime}} & \textit{Technique} &  $F(2,386)=73.9$ & ***&0.28\dag \\
\cline{3-6}
 &  & \textit{Vision} &  $F(1,34)=0.3$ & 0.584&--\\
\cline{3-6}
 &  & \textit{Size} &  $F(1,386)=81.5$ & ***&0.17\dag \\
\cline{3-6}
 &  & \textit{Technique} x \textit{Vision} &  $F(2,386)=10.3$ & ***&0.05 \\
\cline{3-6}
 &  & \textit{Technique} x \textit{Size} &  $F(2,386)=15.5$ & ***&0.07\dag \\
\cline{3-6}
 &  & \textit{Vision} x \textit{Size} &  $F(1,386)=1.1$ & 0.305&--\\
\cline{3-6}
 &  & \textit{Technique} x \textit{Vision} x \textit{Size} &  $F(2,386)=10.7$ & ***&0.05\\
\cline{2-6}
& \multirow{7}{*}{\textit{ReentryTimes}} & \textit{Technique} &  $F(2,386)=86.6$ & ***&0.31\dag \\
\cline{3-6}
 &  & \textit{Vision} &  $F(1,34)=2.0$ & 0.168 & -- \\
\cline{3-6}
 &  & \textit{Size} &  $F(1,386)=97.1$ & ***&0.20\dag \\
\cline{3-6}
 &  & \textit{Technique} x \textit{Vision} &  $F(2,386)=10.6$ & ***&0.05\\
\cline{3-6}
 &  & \textit{Technique} x \textit{Size} &  $F(2,386)=17.1$ & ***&0.08\dag \\
\cline{3-6}
 &  & \textit{Vision} x \textit{Size} &  $F(1,386)=0.1$ & 0.824&-- \\
\cline{3-6}
 &  & \textit{Technique} x \textit{Vision} x \textit{Size} &  $F(2,386)=5.3$ & **&0.03\\
 \cline{2-6}
& \multirow{7}{*}{\textit{WalkingTime}} & \textit{Technique} &  $F(2,386)=54.9$ & ***&0.22\dag \\
\cline{3-6}
 &  & \textit{Vision} &  $F(1,34)=15.4$ & *** & 0.31\dag  \\
\cline{3-6}
 &  & \textit{Size} &  $F(1,386)=146.6$ & ***&0.28\dag \\
\cline{3-6}
 &  & \textit{Technique} x \textit{Vision} &  $F(2,386)=1.2$ & 0.295& -- \\
\cline{3-6}
 &  & \textit{Technique} x \textit{Size} &  $F(2,386)=13.9$ & ***&0.07\dag \\
\cline{3-6}
 &  & \textit{Vision} x \textit{Size} &  $F(1,386)=3.4$ & 0.066& <0.01 \\
\cline{3-6}
 &  & \textit{Technique} x \textit{Vision} x \textit{Size} &  $F(2,386)=0.2$ & 0.820&-- \\
 \cline{2-6}
& \multirow{7}{*}{\textit{SelectionTiming}} & \textit{Technique} &  $F(2,386)=101.9$ & ***&0.35\dag \\
\cline{3-6}
 &  & \textit{Vision} &  $F(1,34)=2.7$ & 0.108 & -- \\
\cline{3-6}
 &  & \textit{Size} &  $F(1,386)=105.8$ & ***&0.22\dag \\
\cline{3-6}
 &  & \textit{Technique} x \textit{Vision} &  $F(2,386)=5.0$ & **&0.03\\
\cline{3-6}
 & \textit{InWalking} & \textit{Technique} x \textit{Size} &  $F(2,386)=5.3$ & **&0.03\\
\cline{3-6}
 &  & \textit{Vision} x \textit{Size} &  $F(1,386)=0.5$ & 0.474& --\\
\cline{3-6}
 &  & \textit{Technique} x \textit{Vision} x \textit{Size} &  $F(2,386)=4.8$ & **&0.02\\
\Xhline{2\arrayrulewidth}
\end{tabular}
\caption{Statistical results for comparing selection behavior between low vision and sighted participants.
Significant codes: *** $p < .001$, ** $p < .01$, * $p < .05$. \dag indicates effect size reached at least a medium level ($\eta_p^2 \geq 0.06$). }
\label{tab:stats_sightedandPLV}
\end{table}

\clearpage

\section{Statistical Effects of Visual Ability on Selection Behavior} \label{sec:stats_va_fov}

\begin{table}[H]
\centering
\scriptsize
\begin{tabular}{>{\centering\arraybackslash}m{1cm} >{\centering\arraybackslash}m{2cm} >{\centering\arraybackslash}m{3.5cm} > {\centering\arraybackslash}m{3cm} > {\centering\arraybackslash}m{1cm}
}
\Xhline{2\arrayrulewidth}
\textbf{Task} & \textbf{Measure} & \textbf{Parameter} & \textbf{F} & \textbf{P-value} \\
\Xhline{2\arrayrulewidth}
\multirow{36}{*}{Static} & \multirow{9}{*}{\textit{SelectionTime}} & \textit{(Intercept)} &  $\chi^2(1)=239.0$ & ***\\
\cline{3-5}
 &  &  \textit{Technique} &  $\chi^2(2)=125.5$ & ***\\
\cline{3-5}
 &  & \textit{VisualAcuity} &  $\chi^2(1)=6.8$ & ** \\
\cline{3-5}
 &  & \textit{PeripheralVision} &  $\chi^2(1)=6.0$ & * \\
\cline{3-5}
 &  & \textit{Technique} x \textit{VisualAcuity} &  $\chi^2(2)=1.4$&0.495 \\
\cline{3-5}
 &  & \textit{Technique} x \textit{PeripheralVision} &  $\chi^2(2)=4.3$ & 0.118 \\
\cline{3-5}
 &  & \textit{VisualAcuity} x \textit{PeripheralVision} &  $\chi^2(1)=5.7$ & * \\
\cline{3-5}
 &  & \textit{Technique} x \textit{VisualAcuity} x \textit{PeripheralVision} &  $\chi^2(2)=9.0$ & * \\
\cline{2-5}
& \multirow{9}{*}{\textit{PointingTime}} & \textit{(Intercept)} &  $\chi^2(1)=0.5$ & 0.471\\
\cline{3-5}
 &  &  \textit{Technique} &  $\chi^2(2)=151.7$ & ***\\
\cline{3-5}
 &  & \textit{VisualAcuity} &  $\chi^2(1)=3.1$ & 0.079\\
\cline{3-5}
 &  & \textit{PeripheralVision} &  $\chi^2(1)=2.8$ & 0.094 \\
\cline{3-5}
 &  & \textit{Technique} x \textit{VisualAcuity} &  $\chi^2(2)=5.6$ & 0.062 \\
\cline{3-5}
 &  & \textit{Technique} x \textit{PeripheralVision} &  $\chi^2(2)=8.2$ & * \\
\cline{3-5}
 &  & \textit{VisualAcuity} x \textit{PeripheralVision} &  $\chi^2(1)=3.3$ & 0.071 \\
\cline{3-5}
 &  & \textit{Technique} x \textit{VisualAcuity} x \textit{PeripheralVision} &  $\chi^2(2)=15.9$ & *** \\
\cline{2-5}
& \multirow{9}{*}{\textit{ConfirmationTime}} & \textit{(Intercept)} &  $\chi^2(1)=3.4$ & 0.065\\
\cline{3-5}
 &  &  \textit{Technique} &  $\chi^2(2)=28.5$ & ***\\
\cline{3-5}
 &  & \textit{VisualAcuity} &  $\chi^2(1)=5.2$ & *\\
\cline{3-5}
 &  & \textit{PeripheralVision} &  $\chi^2(1)=3.3$ & 0.071 \\
\cline{3-5}
 &  & \textit{Technique} x \textit{VisualAcuity} &  $\chi^2(2)=3.8$ & 0.150\\
\cline{3-5}
 &  & \textit{Technique} x \textit{PeripheralVision} &  $\chi^2(2)=0.6$ & 0.747 \\
\cline{3-5}
 &  & \textit{VisualAcuity} x \textit{PeripheralVision} &  $\chi^2(1)=3.2$ & 0.074\\
\cline{3-5}
 &  & \textit{Technique} x \textit{VisualAcuity} x \textit{PeripheralVision} &  $\chi^2(2)=6.8$ & * \\
\cline{2-5}
& \multirow{9}{*}{\textit{ReentryTimes}} & \textit{(Intercept)} &  $\chi^2(1)=57.7$ & ***\\
\cline{3-5}
 &  &  \textit{Technique} &  $\chi^2(2)=51.1$ & ***\\
\cline{3-5}
 &  & \textit{VisualAcuity} &  $\chi^2(1)=2.9$ & 0.089\\
\cline{3-5}
 &  & \textit{PeripheralVision} &  $\chi^2(1)=1.6$ & 0.208\\
\cline{3-5}
 &  & \textit{Technique} x \textit{VisualAcuity} &  $\chi^2(2)=1.2$ & 0.550\\
\cline{3-5}
 &  & \textit{Technique} x \textit{PeripheralVision} &  $\chi^2(2)=1.8$ & 0.406 \\
\cline{3-5}
 &  & \textit{VisualAcuity} x \textit{PeripheralVision} &  $\chi^2(1)=5.0$ & *\\
\cline{3-5}
 &  & \textit{Technique} x \textit{VisualAcuity} x \textit{PeripheralVision} &  $\chi^2(2)=5.5$ & 0.065 \\
\hline

\multirow{36}{*}{Dynamic} & \multirow{9}{*}{\textit{SelectionTime}} & \textit{(Intercept)} &  $\chi^2(1)=428.1$ & ***\\
\cline{3-5}
 &  &  \textit{Technique} &  $\chi^2(2)=34.5$ & ***\\
\cline{3-5}
 &  & \textit{VisualAcuity} &  $\chi^2(1)=11.9$ & ***\\
\cline{3-5}
 &  & \textit{PeripheralVision} &  $\chi^2(1)=22.6$ & ***\\
\cline{3-5}
 &  & \textit{Technique} x \textit{VisualAcuity} &  $\chi^2(2)=6.8$ &*\\
\cline{3-5}
 &  & \textit{Technique} x \textit{PeripheralVision} &  $\chi^2(2)=2.4$ & 0.307 \\
\cline{3-5}
 &  & \textit{VisualAcuity} x \textit{PeripheralVision} &  $\chi^2(1)=11.7$ & *** \\
\cline{3-5}
 &  & \textit{Technique} x \textit{VisualAcuity} x \textit{PeripheralVision} &  $\chi^2(2)=9.0$ & *\\
\cline{2-5}
& \multirow{9}{*}{\textit{PointingTime}} & \textit{(Intercept)} &  $\chi^2(1)=83.1$ & ***\\
\cline{3-5}
 &  &  \textit{Technique} &  $\chi^2(2)=10.5$ & **\\
\cline{3-5}
 &  & \textit{VisualAcuity} &  $\chi^2(1)=7.0$ & ** \\
\cline{3-5}
 &  & \textit{PeripheralVision} &  $\chi^2(1)=16.7$ & *** \\
\cline{3-5}
 &  & \textit{Technique} x \textit{VisualAcuity} &  $\chi^2(2)=4.2$ & 0.133 \\
\cline{3-5}
 &  & \textit{Technique} x \textit{PeripheralVision} &  $\chi^2(2)=1.4$ & 0.496\\
\cline{3-5}
 &  & \textit{VisualAcuity} x \textit{PeripheralVision} &  $\chi^2(1)=11.3$ & ***\\
\cline{3-5}
 &  & \textit{Technique} x \textit{VisualAcuity} x \textit{PeripheralVision} &  $\chi^2(2)=3.3$ & 0.194 \\
\cline{2-5}
& \multirow{9}{*}{\textit{ConfirmationTime}} & \textit{(Intercept)} &  $\chi^2(1)=52.7$ & ***\\
\cline{3-5}
 &  &  \textit{Technique} &  $\chi^2(2)=42.1$ & ***\\
\cline{3-5}
 &  & \textit{VisualAcuity} &  $\chi^2(1)=8.2$ & ** \\
\cline{3-5}
 &  & \textit{PeripheralVision} &  $\chi^2(1)=6.5$ & * \\
\cline{3-5}
 &  & \textit{Technique} x \textit{VisualAcuity} &  $\chi^2(2)=1.5$ & 0.479 \\
\cline{3-5}
 &  & \textit{Technique} x \textit{PeripheralVision} &  $\chi^2(2)=1.0$ & 0.607 \\
\cline{3-5}
 &  & \textit{VisualAcuity} x \textit{PeripheralVision} &  $\chi^2(1)=2.6$ & 0.109 \\
\cline{3-5}
 &  & \textit{Technique} x \textit{VisualAcuity} x \textit{PeripheralVision} &  $\chi^2(2)=1.9$ & 0.378\\
\cline{2-5}
& \multirow{9}{*}{\textit{ReentryTimes}} & \textit{(Intercept)} &  $\chi^2(1)=1.7$ & 0.193\\
\cline{3-5}
 &  &  \textit{Technique} &  $\chi^2(2)=42.2$ & ***\\
\cline{3-5}
 &  & \textit{VisualAcuity} &  $\chi^2(1)=7.5$ & ** \\
\cline{3-5}
 &  & \textit{PeripheralVision} &  $\chi^2(1)=1.9$ & 0.167\\
\cline{3-5}
 &  & \textit{Technique} x \textit{VisualAcuity} &  $\chi^2(2)=11.1$ & **\\
\cline{3-5}
 &  & \textit{Technique} x \textit{PeripheralVision} &  $\chi^2(2)=1.5$ & 0.466\\
\cline{3-5}
 &  & \textit{VisualAcuity} x \textit{PeripheralVision} &  $\chi^2(1)=0.6$ & 0.436\\
\cline{3-5}
 &  & \textit{Technique} x \textit{VisualAcuity} x \textit{PeripheralVision} &  $\chi^2(2)=1.0$ & 0.621\\
\Xhline{2\arrayrulewidth}
\end{tabular}
\caption{Statistical results for comparing selection behavior among low vision participants with different visual abilities.
Significant codes: *** $p < .001$, ** $p < .01$, * $p < .05$.}
\label{tab:stats_va_fov}
\end{table}

%% file: main.bbl

\begin{thebibliography}{104}


\ifx \showCODEN    \undefined \def \showCODEN     #1{\unskip}     \fi
\ifx \showISBNx    \undefined \def \showISBNx     #1{\unskip}     \fi
\ifx \showISBNxiii \undefined \def \showISBNxiii  #1{\unskip}     \fi
\ifx \showISSN     \undefined \def \showISSN      #1{\unskip}     \fi
\ifx \showLCCN     \undefined \def \showLCCN      #1{\unskip}     \fi
\ifx \shownote     \undefined \def \shownote      #1{#1}          \fi
\ifx \showarticletitle \undefined \def \showarticletitle #1{#1}   \fi
\ifx \showURL      \undefined \def \showURL       {\relax}        \fi
\providecommand\bibfield[2]{#2}
\providecommand\bibinfo[2]{#2}
\providecommand\natexlab[1]{#1}
\providecommand\showeprint[2][]{arXiv:#2}

\bibitem[2026(2026)]%
        {microsoftHeadgazeCommit}
\bibfield{author}{\bibinfo{person}{Microsoft~Build 2026}.} \bibinfo{year}{2026}\natexlab{}.
\newblock \bibinfo{title}{Head-gaze and commit - Mixed Reality --- learn.microsoft.com}.
\newblock \bibinfo{howpublished}{\url{https://learn.microsoft.com/en-us/windows/mixed-reality/design/gaze-and-commit-head}}.
\newblock
\newblock
\shownote{[Accessed 17-03-2026]}.


\bibitem[Abbott et~al\mbox{.}(2024)]%
        {abbott2024prevalence}
\bibfield{author}{\bibinfo{person}{Kaleb Abbott}, \bibinfo{person}{Kara~S Hanson}, {and} \bibinfo{person}{James Lally}.} \bibinfo{year}{2024}\natexlab{}.
\newblock \showarticletitle{Prevalence of dry eye disease in the low vision population at the University of Colorado}.
\newblock \bibinfo{journal}{\emph{Journal of Optometry}} \bibinfo{volume}{17}, \bibinfo{number}{2} (\bibinfo{year}{2024}), \bibinfo{pages}{100501}.
\newblock


\bibitem[Abe et~al\mbox{.}(2025)]%
        {abe2025can}
\bibfield{author}{\bibinfo{person}{Yuki Abe}, \bibinfo{person}{Keisuke Matsushima}, \bibinfo{person}{Kotaro Hara}, \bibinfo{person}{Daisuke Sakamoto}, {and} \bibinfo{person}{Tetsuo Ono}.} \bibinfo{year}{2025}\natexlab{}.
\newblock \showarticletitle{“I can run at night!": Using Augmented Reality to Support Nighttime Guided Running for Low-vision Runners}. In \bibinfo{booktitle}{\emph{Proceedings of the 2025 CHI Conference on Human Factors in Computing Systems}}. \bibinfo{pages}{1--20}.
\newblock


\bibitem[Angelopoulos et~al\mbox{.}(2019)]%
        {angelopoulos2019enhanced}
\bibfield{author}{\bibinfo{person}{Anastasios~Nikolas Angelopoulos}, \bibinfo{person}{Hossein Ameri}, \bibinfo{person}{Debbie Mitra}, {and} \bibinfo{person}{Mark Humayun}.} \bibinfo{year}{2019}\natexlab{}.
\newblock \showarticletitle{Enhanced depth navigation through augmented reality depth mapping in patients with low vision}.
\newblock \bibinfo{journal}{\emph{Scientific reports}} \bibinfo{volume}{9}, \bibinfo{number}{1} (\bibinfo{year}{2019}), \bibinfo{pages}{11230}.
\newblock


\bibitem[B{\^a}ce et~al\mbox{.}(2016)]%
        {bace2016ubigaze}
\bibfield{author}{\bibinfo{person}{Mihai B{\^a}ce}, \bibinfo{person}{Teemu Lepp{\"a}nen}, \bibinfo{person}{David~Gil De~Gomez}, {and} \bibinfo{person}{Argenis~Ramirez Gomez}.} \bibinfo{year}{2016}\natexlab{}.
\newblock \showarticletitle{ubiGaze: ubiquitous augmented reality messaging using gaze gestures}.
\newblock In \bibinfo{booktitle}{\emph{SIGGRAPH ASIA 2016 Mobile Graphics and Interactive Applications}}. \bibinfo{pages}{1--5}.
\newblock


\bibitem[Baumann and Dierkes(2023)]%
        {baumann2023neon}
\bibfield{author}{\bibinfo{person}{Chris Baumann} {and} \bibinfo{person}{Kai Dierkes}.} \bibinfo{year}{2023}\natexlab{}.
\newblock \showarticletitle{Neon accuracy test report}.
\newblock \bibinfo{journal}{\emph{Pupil Labs}}  \bibinfo{volume}{10} (\bibinfo{year}{2023}).
\newblock


\bibitem[Bernardos et~al\mbox{.}(2016)]%
        {bernardos2016comparison}
\bibfield{author}{\bibinfo{person}{Ana~M Bernardos}, \bibinfo{person}{David G{\'o}mez}, {and} \bibinfo{person}{Jos{\'e}~R Casar}.} \bibinfo{year}{2016}\natexlab{}.
\newblock \showarticletitle{A comparison of head pose and deictic pointing interaction methods for smart environments}.
\newblock \bibinfo{journal}{\emph{International Journal of Human-Computer Interaction}} \bibinfo{volume}{32}, \bibinfo{number}{4} (\bibinfo{year}{2016}), \bibinfo{pages}{325--351}.
\newblock


\bibitem[Blattgerste et~al\mbox{.}(2018)]%
        {blattgerste2018advantages}
\bibfield{author}{\bibinfo{person}{Jonas Blattgerste}, \bibinfo{person}{Patrick Renner}, {and} \bibinfo{person}{Thies Pfeiffer}.} \bibinfo{year}{2018}\natexlab{}.
\newblock \showarticletitle{Advantages of eye-gaze over head-gaze-based selection in virtual and augmented reality under varying field of views}. In \bibinfo{booktitle}{\emph{Proceedings of the workshop on communication by gaze interaction}}. \bibinfo{pages}{1--9}.
\newblock


\bibitem[Braun and Clarke(2006)]%
        {braun2006using}
\bibfield{author}{\bibinfo{person}{Virginia Braun} {and} \bibinfo{person}{Victoria Clarke}.} \bibinfo{year}{2006}\natexlab{}.
\newblock \showarticletitle{Using thematic analysis in psychology}.
\newblock \bibinfo{journal}{\emph{Qualitative research in psychology}} \bibinfo{volume}{3}, \bibinfo{number}{2} (\bibinfo{year}{2006}), \bibinfo{pages}{77--101}.
\newblock


\bibitem[Breslow and Clayton(1993)]%
        {breslow1993approximate}
\bibfield{author}{\bibinfo{person}{Norman~E Breslow} {and} \bibinfo{person}{David~G Clayton}.} \bibinfo{year}{1993}\natexlab{}.
\newblock \showarticletitle{Approximate inference in generalized linear mixed models}.
\newblock \bibinfo{journal}{\emph{Journal of the American statistical Association}} \bibinfo{volume}{88}, \bibinfo{number}{421} (\bibinfo{year}{1993}), \bibinfo{pages}{9--25}.
\newblock


\bibitem[Chen et~al\mbox{.}(2023)]%
        {chen2023gazeraycursor}
\bibfield{author}{\bibinfo{person}{Di~Laura Chen}, \bibinfo{person}{Marcello Giordano}, \bibinfo{person}{Hrvoje Benko}, \bibinfo{person}{Tovi Grossman}, {and} \bibinfo{person}{Stephanie Santosa}.} \bibinfo{year}{2023}\natexlab{}.
\newblock \showarticletitle{Gazeraycursor: Facilitating virtual reality target selection by blending gaze and controller raycasting}. In \bibinfo{booktitle}{\emph{Proceedings of the 29th ACM Symposium on Virtual Reality Software and Technology}}. \bibinfo{pages}{1--11}.
\newblock


\bibitem[Chen et~al\mbox{.}(2025)]%
        {chen2025visimark}
\bibfield{author}{\bibinfo{person}{Ruijia Chen}, \bibinfo{person}{Junru Jiang}, \bibinfo{person}{Pragati Maheshwary}, \bibinfo{person}{Brianna~R Cochran}, {and} \bibinfo{person}{Yuhang Zhao}.} \bibinfo{year}{2025}\natexlab{}.
\newblock \showarticletitle{VisiMark: Characterizing and Augmenting Landmarks for People with Low Vision in Augmented Reality to Support Indoor Navigation}. In \bibinfo{booktitle}{\emph{Proceedings of the 2025 CHI Conference on Human Factors in Computing Systems}}. \bibinfo{pages}{1--20}.
\newblock


\bibitem[Chen et~al\mbox{.}(2021)]%
        {chen2021adaptive}
\bibfield{author}{\bibinfo{person}{Xiuli Chen}, \bibinfo{person}{Aditya Acharya}, \bibinfo{person}{Antti Oulasvirta}, {and} \bibinfo{person}{Andrew Howes}.} \bibinfo{year}{2021}\natexlab{}.
\newblock \showarticletitle{An adaptive model of gaze-based selection}. In \bibinfo{booktitle}{\emph{Proceedings of the 2021 CHI Conference on Human Factors in Computing Systems}}. \bibinfo{pages}{1--11}.
\newblock


\bibitem[Clarke and Braun(2017)]%
        {clarke2017thematic}
\bibfield{author}{\bibinfo{person}{Victoria Clarke} {and} \bibinfo{person}{Virginia Braun}.} \bibinfo{year}{2017}\natexlab{}.
\newblock \showarticletitle{Thematic analysis}.
\newblock \bibinfo{journal}{\emph{The journal of positive psychology}} \bibinfo{volume}{12}, \bibinfo{number}{3} (\bibinfo{year}{2017}), \bibinfo{pages}{297--298}.
\newblock


\bibitem[Cohen(2013)]%
        {cohen2013statistical}
\bibfield{author}{\bibinfo{person}{Jacob Cohen}.} \bibinfo{year}{2013}\natexlab{}.
\newblock \bibinfo{booktitle}{\emph{Statistical power analysis for the behavioral sciences}}.
\newblock \bibinfo{publisher}{routledge}.
\newblock


\bibitem[Cournia et~al\mbox{.}(2003)]%
        {cournia2003gaze}
\bibfield{author}{\bibinfo{person}{Nathan Cournia}, \bibinfo{person}{John~D Smith}, {and} \bibinfo{person}{Andrew~T Duchowski}.} \bibinfo{year}{2003}\natexlab{}.
\newblock \showarticletitle{Gaze-vs. hand-based pointing in virtual environments}. In \bibinfo{booktitle}{\emph{CHI'03 extended abstracts on Human factors in computing systems}}. \bibinfo{pages}{772--773}.
\newblock


\bibitem[Crossland et~al\mbox{.}(2005)]%
        {crossland2005preferred}
\bibfield{author}{\bibinfo{person}{Michael~D Crossland}, \bibinfo{person}{Louise~E Culham}, \bibinfo{person}{Stamatina~A Kabanarou}, {and} \bibinfo{person}{Gary~S Rubin}.} \bibinfo{year}{2005}\natexlab{}.
\newblock \showarticletitle{Preferred retinal locus development in patients with macular disease}.
\newblock \bibinfo{journal}{\emph{Ophthalmology}} \bibinfo{volume}{112}, \bibinfo{number}{9} (\bibinfo{year}{2005}), \bibinfo{pages}{1579--1585}.
\newblock


\bibitem[Drewes et~al\mbox{.}(2019)]%
        {drewes2019dialplates}
\bibfield{author}{\bibinfo{person}{Heiko Drewes}, \bibinfo{person}{Mohamed Khamis}, {and} \bibinfo{person}{Florian Alt}.} \bibinfo{year}{2019}\natexlab{}.
\newblock \showarticletitle{Dialplates: Enabling pursuits-based user interfaces with large target numbers}. In \bibinfo{booktitle}{\emph{Proceedings of the 18th International Conference on Mobile and Ubiquitous Multimedia}}. \bibinfo{pages}{1--10}.
\newblock


\bibitem[Esteves et~al\mbox{.}(2020)]%
        {esteves2020comparing}
\bibfield{author}{\bibinfo{person}{Augusto Esteves}, \bibinfo{person}{Yonghwan Shin}, {and} \bibinfo{person}{Ian Oakley}.} \bibinfo{year}{2020}\natexlab{}.
\newblock \showarticletitle{Comparing selection mechanisms for gaze input techniques in head-mounted displays}.
\newblock \bibinfo{journal}{\emph{International Journal of Human-Computer Studies}}  \bibinfo{volume}{139} (\bibinfo{year}{2020}), \bibinfo{pages}{102414}.
\newblock


\bibitem[Esteves et~al\mbox{.}(2015)]%
        {esteves2015orbits}
\bibfield{author}{\bibinfo{person}{Augusto Esteves}, \bibinfo{person}{Eduardo Velloso}, \bibinfo{person}{Andreas Bulling}, {and} \bibinfo{person}{Hans Gellersen}.} \bibinfo{year}{2015}\natexlab{}.
\newblock \showarticletitle{Orbits: Gaze interaction for smart watches using smooth pursuit eye movements}. In \bibinfo{booktitle}{\emph{Proceedings of the 28th annual ACM symposium on user interface software \& technology}}. \bibinfo{pages}{457--466}.
\newblock


\bibitem[Fashimpaur et~al\mbox{.}(2020)]%
        {fashimpaur2020pinchtype}
\bibfield{author}{\bibinfo{person}{Jacqui Fashimpaur}, \bibinfo{person}{Kenrick Kin}, {and} \bibinfo{person}{Matt Longest}.} \bibinfo{year}{2020}\natexlab{}.
\newblock \showarticletitle{Pinchtype: Text entry for virtual and augmented reality using comfortable thumb to fingertip pinches}. In \bibinfo{booktitle}{\emph{Extended abstracts of the 2020 CHI conference on human factors in computing systems}}. \bibinfo{pages}{1--7}.
\newblock


\bibitem[{Federal Highway Administration}(2003)]%
        {FHWA_MUTCD_Part4E_2003}
\bibfield{author}{\bibinfo{person}{{Federal Highway Administration}}.} \bibinfo{year}{2003}\natexlab{}.
\newblock \bibinfo{title}{{Manual on Uniform Traffic Control Devices (MUTCD), Part 4E: Pedestrian Control Features}}.
\newblock \bibinfo{howpublished}{\url{https://mutcd.fhwa.dot.gov/HTM/2003/part4/part4e.htm}}.
\newblock
\newblock
\shownote{U.S. Department of Transportation}.


\bibitem[Ferris~III et~al\mbox{.}(1982)]%
        {ferris1982new}
\bibfield{author}{\bibinfo{person}{Frederick~L Ferris~III}, \bibinfo{person}{Aaron Kassoff}, \bibinfo{person}{George~H Bresnick}, {and} \bibinfo{person}{Ian Bailey}.} \bibinfo{year}{1982}\natexlab{}.
\newblock \showarticletitle{New visual acuity charts for clinical research}.
\newblock \bibinfo{journal}{\emph{American journal of ophthalmology}} \bibinfo{volume}{94}, \bibinfo{number}{1} (\bibinfo{year}{1982}), \bibinfo{pages}{91--96}.
\newblock


\bibitem[for~the Blind(2026)]%
        {afbVisionLegal}
\bibfield{author}{\bibinfo{person}{American~Foundation for~the Blind}.} \bibinfo{year}{2026}\natexlab{}.
\newblock \bibinfo{title}{Low Vision and Legal Blindness Terms and Descriptions --- afb.org}.
\newblock \bibinfo{howpublished}{\url{https://afb.org/blindness-and-low-vision/eye-conditions/low-vision-and-legal-blindness-terms-and-descriptions}}.
\newblock
\newblock
\shownote{[Accessed 27-03-2026]}.


\bibitem[Fox et~al\mbox{.}(2023)]%
        {fox2023using}
\bibfield{author}{\bibinfo{person}{Dylan~R Fox}, \bibinfo{person}{Ahmad Ahmadzada}, \bibinfo{person}{Clara~T Friedman}, \bibinfo{person}{Shiri Azenkot}, \bibinfo{person}{Marlena~A Chu}, \bibinfo{person}{Roberto Manduchi}, {and} \bibinfo{person}{Emily~A Cooper}.} \bibinfo{year}{2023}\natexlab{}.
\newblock \showarticletitle{Using augmented reality to cue obstacles for people with low vision}.
\newblock \bibinfo{journal}{\emph{Optics Express}} \bibinfo{volume}{31}, \bibinfo{number}{4} (\bibinfo{year}{2023}), \bibinfo{pages}{6827--6848}.
\newblock


\bibitem[Ga{\l}ecki and Burzykowski(2012)]%
        {galecki2012linear}
\bibfield{author}{\bibinfo{person}{Andrzej Ga{\l}ecki} {and} \bibinfo{person}{Tomasz Burzykowski}.} \bibinfo{year}{2012}\natexlab{}.
\newblock \showarticletitle{Linear mixed-effects model}.
\newblock In \bibinfo{booktitle}{\emph{Linear mixed-effects models using R: a step-by-step approach}}. \bibinfo{publisher}{Springer}, \bibinfo{pages}{245--273}.
\newblock


\bibitem[Gavgiotaki et~al\mbox{.}(2023)]%
        {gavgiotaki2023gesture}
\bibfield{author}{\bibinfo{person}{Despoina Gavgiotaki}, \bibinfo{person}{Stavroula Ntoa}, \bibinfo{person}{George Margetis}, \bibinfo{person}{Konstantinos~C Apostolakis}, {and} \bibinfo{person}{Constantine Stephanidis}.} \bibinfo{year}{2023}\natexlab{}.
\newblock \showarticletitle{Gesture-based interaction for AR systems: a short review}. In \bibinfo{booktitle}{\emph{Proceedings of the 16th International Conference on PErvasive Technologies Related to Assistive Environments}}. \bibinfo{pages}{284--292}.
\newblock


\bibitem[Hart and Staveland(1988)]%
        {hart1988development}
\bibfield{author}{\bibinfo{person}{Sandra~G Hart} {and} \bibinfo{person}{Lowell~E Staveland}.} \bibinfo{year}{1988}\natexlab{}.
\newblock \showarticletitle{Development of NASA-TLX (Task Load Index): Results of empirical and theoretical research}.
\newblock In \bibinfo{booktitle}{\emph{Advances in psychology}}. Vol.~\bibinfo{volume}{52}. \bibinfo{publisher}{Elsevier}, \bibinfo{pages}{139--183}.
\newblock


\bibitem[Herskovitz et~al\mbox{.}(2023)]%
        {herskovitz2023hacking}
\bibfield{author}{\bibinfo{person}{Jaylin Herskovitz}, \bibinfo{person}{Andi Xu}, \bibinfo{person}{Rahaf Alharbi}, {and} \bibinfo{person}{Anhong Guo}.} \bibinfo{year}{2023}\natexlab{}.
\newblock \showarticletitle{Hacking, switching, combining: understanding and supporting DIY assistive technology design by blind people}. In \bibinfo{booktitle}{\emph{Proceedings of the 2023 CHI conference on human factors in computing systems}}. \bibinfo{pages}{1--17}.
\newblock


\bibitem[Horizon(2026)]%
        {metaHead}
\bibfield{author}{\bibinfo{person}{Meta Horizon}.} \bibinfo{year}{2026}\natexlab{}.
\newblock \bibinfo{title}{Head --- developers.meta.com}.
\newblock \bibinfo{howpublished}{\url{https://developers.meta.com/horizon/design/head}}.
\newblock
\newblock
\shownote{[Accessed 17-03-2026]}.


\bibitem[Huang et~al\mbox{.}(2019)]%
        {huang2019augmented}
\bibfield{author}{\bibinfo{person}{Jonathan Huang}, \bibinfo{person}{Max Kinateder}, \bibinfo{person}{Matt~J Dunn}, \bibinfo{person}{Wojciech Jarosz}, \bibinfo{person}{Xing-Dong Yang}, {and} \bibinfo{person}{Emily~A Cooper}.} \bibinfo{year}{2019}\natexlab{}.
\newblock \showarticletitle{An augmented reality sign-reading assistant for users with reduced vision}.
\newblock \bibinfo{journal}{\emph{PloS one}} \bibinfo{volume}{14}, \bibinfo{number}{1} (\bibinfo{year}{2019}), \bibinfo{pages}{e0210630}.
\newblock


\bibitem[Hwang and Peli(2014)]%
        {hwang2014augmented}
\bibfield{author}{\bibinfo{person}{Alex~D Hwang} {and} \bibinfo{person}{Eli Peli}.} \bibinfo{year}{2014}\natexlab{}.
\newblock \showarticletitle{An augmented-reality edge enhancement application for Google Glass}.
\newblock \bibinfo{journal}{\emph{Optometry and vision science}} \bibinfo{volume}{91}, \bibinfo{number}{8} (\bibinfo{year}{2014}), \bibinfo{pages}{1021--1030}.
\newblock


\bibitem[Hyrskykari et~al\mbox{.}(2012)]%
        {hyrskykari2012gaze}
\bibfield{author}{\bibinfo{person}{Aulikki Hyrskykari}, \bibinfo{person}{Howell Istance}, {and} \bibinfo{person}{Stephen Vickers}.} \bibinfo{year}{2012}\natexlab{}.
\newblock \showarticletitle{Gaze gestures or dwell-based interaction?}. In \bibinfo{booktitle}{\emph{Proceedings of the Symposium on Eye Tracking Research and Applications}}. \bibinfo{pages}{229--232}.
\newblock


\bibitem[Inc.(2026a)]%
        {appleMagnifyDescribe}
\bibfield{author}{\bibinfo{person}{Apple Inc.}} \bibinfo{year}{2026}\natexlab{a}.
\newblock \bibinfo{title}{Magnify or describe things around you with Magnifier on iPhone --- support.apple.com}.
\newblock \bibinfo{howpublished}{\url{https://support.apple.com/guide/iphone/magnify-or-describe-things-around-you-iphe867dc99c/ios}}.
\newblock
\newblock
\shownote{[Accessed 17-03-2026]}.


\bibitem[Inc.(2026b)]%
        {applePointer}
\bibfield{author}{\bibinfo{person}{Apple Inc.}} \bibinfo{year}{2026}\natexlab{b}.
\newblock \bibinfo{title}{Use a pointer to navigate your Apple Vision Pro}.
\newblock \bibinfo{howpublished}{\url{https://support.apple.com/guide/apple-vision-pro/use-a-pointer-to-navigate-tan3869c8a85/visionos}}.
\newblock
\newblock
\shownote{[Accessed 17-03-2026]}.


\bibitem[Jacob(1990)]%
        {jacob1990you}
\bibfield{author}{\bibinfo{person}{Robert~JK Jacob}.} \bibinfo{year}{1990}\natexlab{}.
\newblock \showarticletitle{What you look at is what you get: eye movement-based interaction techniques}. In \bibinfo{booktitle}{\emph{Proceedings of the SIGCHI conference on Human factors in computing systems}}. \bibinfo{pages}{11--18}.
\newblock


\bibitem[Jalaliniya et~al\mbox{.}(2014)]%
        {jalaliniya2014head}
\bibfield{author}{\bibinfo{person}{Shahram Jalaliniya}, \bibinfo{person}{Diako Mardanbeigi}, \bibinfo{person}{Thomas Pederson}, {and} \bibinfo{person}{Dan~Witzner Hansen}.} \bibinfo{year}{2014}\natexlab{}.
\newblock \showarticletitle{Head and eye movement as pointing modalities for eyewear computers}. In \bibinfo{booktitle}{\emph{2014 11th International Conference on Wearable and Implantable Body Sensor Networks Workshops}}. IEEE, \bibinfo{pages}{50--53}.
\newblock


\bibitem[Jeong et~al\mbox{.}(2023)]%
        {jeong2023gazehand}
\bibfield{author}{\bibinfo{person}{Jaejoon Jeong}, \bibinfo{person}{Soo-Hyung Kim}, \bibinfo{person}{Hyung-Jeong Yang}, \bibinfo{person}{Gun~A Lee}, {and} \bibinfo{person}{Seungwon Kim}.} \bibinfo{year}{2023}\natexlab{}.
\newblock \showarticletitle{GazeHand: A gaze-driven virtual hand interface}.
\newblock \bibinfo{journal}{\emph{IEEE Access}}  \bibinfo{volume}{11} (\bibinfo{year}{2023}), \bibinfo{pages}{133703--133716}.
\newblock


\bibitem[Jiang et~al\mbox{.}(2019)]%
        {jiang2019appearance}
\bibfield{author}{\bibinfo{person}{Jiaqi Jiang}, \bibinfo{person}{Xiaolong Zhou}, \bibinfo{person}{Sixian Chan}, {and} \bibinfo{person}{Shengyong Chen}.} \bibinfo{year}{2019}\natexlab{}.
\newblock \showarticletitle{Appearance-based gaze tracking: A brief review}. In \bibinfo{booktitle}{\emph{International Conference on Intelligent Robotics and Applications}}. Springer, \bibinfo{pages}{629--640}.
\newblock


\bibitem[Jocher et~al\mbox{.}(2023)]%
        {yolov8_ultralytics}
\bibfield{author}{\bibinfo{person}{Glenn Jocher}, \bibinfo{person}{Ayush Chaurasia}, {and} \bibinfo{person}{Jing Qiu}.} \bibinfo{year}{2023}\natexlab{}.
\newblock \bibinfo{booktitle}{\emph{Ultralytics YOLOv8}}.
\newblock
\urldef\tempurl%
\url{https://github.com/ultralytics/ultralytics}
\showURL{%
\tempurl}


\bibitem[Kane et~al\mbox{.}(2009)]%
        {kane2009freedom}
\bibfield{author}{\bibinfo{person}{Shaun~K Kane}, \bibinfo{person}{Chandrika Jayant}, \bibinfo{person}{Jacob~O Wobbrock}, {and} \bibinfo{person}{Richard~E Ladner}.} \bibinfo{year}{2009}\natexlab{}.
\newblock \showarticletitle{Freedom to roam: a study of mobile device adoption and accessibility for people with visual and motor disabilities}. In \bibinfo{booktitle}{\emph{Proceedings of the 11th international ACM SIGACCESS conference on Computers and accessibility}}. \bibinfo{pages}{115--122}.
\newblock


\bibitem[Kaya et~al\mbox{.}(2024)]%
        {kaya2024virtual}
\bibfield{author}{\bibinfo{person}{Cem Kaya}, \bibinfo{person}{Baha~Mert Ersoy}, {and} \bibinfo{person}{Murat Karaca}.} \bibinfo{year}{2024}\natexlab{}.
\newblock \showarticletitle{Virtual reality meets low vision: the development and analysis of MagniVR as an assistive technology}. In \bibinfo{booktitle}{\emph{International Conference on Human-Computer Interaction}}. Springer, \bibinfo{pages}{321--333}.
\newblock


\bibitem[Kerr and Fuad(2019)]%
        {kerr2019real}
\bibfield{author}{\bibinfo{person}{Robert Kerr} {and} \bibinfo{person}{Muhammad Marwan~Muhammad Fuad}.} \bibinfo{year}{2019}\natexlab{}.
\newblock \showarticletitle{A real-time lazy eye correction method for low cost webcams}.
\newblock \bibinfo{journal}{\emph{Procedia Computer Science}}  \bibinfo{volume}{159} (\bibinfo{year}{2019}), \bibinfo{pages}{281--290}.
\newblock


\bibitem[Kim et~al\mbox{.}(2025)]%
        {kim2025pinchcatcher}
\bibfield{author}{\bibinfo{person}{Jinwook Kim}, \bibinfo{person}{Sangmin Park}, \bibinfo{person}{Qiushi Zhou}, \bibinfo{person}{Mar Gonzalez-Franco}, \bibinfo{person}{Jeongmi Lee}, {and} \bibinfo{person}{Ken Pfeuffer}.} \bibinfo{year}{2025}\natexlab{}.
\newblock \showarticletitle{PinchCatcher: Enabling Multi-selection for Gaze+ Pinch}. In \bibinfo{booktitle}{\emph{Proceedings of the 2025 CHI Conference on Human Factors in Computing Systems}}. \bibinfo{pages}{1--16}.
\newblock


\bibitem[Kyt{\"o} et~al\mbox{.}(2018)]%
        {kyto2018pinpointing}
\bibfield{author}{\bibinfo{person}{Mikko Kyt{\"o}}, \bibinfo{person}{Barrett Ens}, \bibinfo{person}{Thammathip Piumsomboon}, \bibinfo{person}{Gun~A Lee}, {and} \bibinfo{person}{Mark Billinghurst}.} \bibinfo{year}{2018}\natexlab{}.
\newblock \showarticletitle{Pinpointing: Precise head-and eye-based target selection for augmented reality}. In \bibinfo{booktitle}{\emph{Proceedings of the 2018 CHI conference on human factors in computing systems}}. \bibinfo{pages}{1--14}.
\newblock


\bibitem[Labs(2026a)]%
        {pupillabsNeonOverview}
\bibfield{author}{\bibinfo{person}{Pupil Labs}.} \bibinfo{year}{2026}\natexlab{a}.
\newblock \bibinfo{title}{Neon - Overview - Eye tracking for research and beyond --- pupil-labs.com}.
\newblock \bibinfo{howpublished}{\url{https://pupil-labs.com/products/neon}}.
\newblock
\newblock
\shownote{[Accessed 19-03-2026]}.


\bibitem[Labs(2026b)]%
        {pupillabsNeonUsing}
\bibfield{author}{\bibinfo{person}{Pupil Labs}.} \bibinfo{year}{2026}\natexlab{b}.
\newblock \bibinfo{title}{Neon - Using Offset Correction to Improve Gaze Accuracy - Pupil Labs Docs --- docs.pupil-labs.com}.
\newblock \bibinfo{howpublished}{\url{https://docs.pupil-labs.com/neon/data-collection/offset-correction/}}.
\newblock
\newblock
\shownote{[Accessed 31-03-2026]}.


\bibitem[Lee et~al\mbox{.}(2024a)]%
        {lee2024towards}
\bibfield{author}{\bibinfo{person}{Jaewook Lee}, \bibinfo{person}{Yang Li}, \bibinfo{person}{Dylan Bunarto}, \bibinfo{person}{Eujean Lee}, \bibinfo{person}{Olivia~H Wang}, \bibinfo{person}{Adrian Rodriguez}, \bibinfo{person}{Yuhang Zhao}, \bibinfo{person}{Yapeng Tian}, {and} \bibinfo{person}{Jon~E Froehlich}.} \bibinfo{year}{2024}\natexlab{a}.
\newblock \showarticletitle{Towards AI-Powered AR for Enhancing Sports Playability for People with Low Vision: An Exploration of ARSports}. In \bibinfo{booktitle}{\emph{2024 IEEE International Symposium on Mixed and Augmented Reality Adjunct (ISMAR-Adjunct)}}. IEEE, \bibinfo{pages}{228--233}.
\newblock


\bibitem[Lee et~al\mbox{.}(2024b)]%
        {lee2024cookar}
\bibfield{author}{\bibinfo{person}{Jaewook Lee}, \bibinfo{person}{Andrew~D Tjahjadi}, \bibinfo{person}{Jiho Kim}, \bibinfo{person}{Junpu Yu}, \bibinfo{person}{Minji Park}, \bibinfo{person}{Jiawen Zhang}, \bibinfo{person}{Jon~E Froehlich}, \bibinfo{person}{Yapeng Tian}, {and} \bibinfo{person}{Yuhang Zhao}.} \bibinfo{year}{2024}\natexlab{b}.
\newblock \showarticletitle{CookAR: Affordance augmentations in wearable AR to support kitchen tool interactions for people with low vision}. In \bibinfo{booktitle}{\emph{Proceedings of the 37th Annual ACM Symposium on User Interface Software and Technology}}. \bibinfo{pages}{1--16}.
\newblock


\bibitem[Lei and Schuchard(1997)]%
        {lei1997using}
\bibfield{author}{\bibinfo{person}{Hong Lei} {and} \bibinfo{person}{Ronald~A Schuchard}.} \bibinfo{year}{1997}\natexlab{}.
\newblock \showarticletitle{Using two preferred retinal loci for different lighting conditions in patients with central scotomas.}
\newblock \bibinfo{journal}{\emph{Investigative ophthalmology \& visual science}} \bibinfo{volume}{38}, \bibinfo{number}{9} (\bibinfo{year}{1997}), \bibinfo{pages}{1812--1818}.
\newblock


\bibitem[Li et~al\mbox{.}(2025)]%
        {li2025effects}
\bibfield{author}{\bibinfo{person}{Chen Li}, \bibinfo{person}{Xia Zhang}, \bibinfo{person}{Wei Li}, {and} \bibinfo{person}{Yafei Liu}.} \bibinfo{year}{2025}\natexlab{}.
\newblock \showarticletitle{Effects of Dwell Time and Area on Task Performance of Gaze-Triggered Eye Control in AR-Aided Aircraft Cockpits}.
\newblock \bibinfo{journal}{\emph{International Journal of Human--Computer Interaction}} (\bibinfo{year}{2025}), \bibinfo{pages}{1--8}.
\newblock


\bibitem[Lin et~al\mbox{.}(2015a)]%
        {lin2015investigation}
\bibfield{author}{\bibinfo{person}{Chiuhsiang~Joe Lin}, \bibinfo{person}{Sui-Hua Ho}, {and} \bibinfo{person}{Yan-Jyun Chen}.} \bibinfo{year}{2015}\natexlab{a}.
\newblock \showarticletitle{An investigation of pointing postures in a 3D stereoscopic environment}.
\newblock \bibinfo{journal}{\emph{Applied ergonomics}}  \bibinfo{volume}{48} (\bibinfo{year}{2015}), \bibinfo{pages}{154--163}.
\newblock


\bibitem[Lin et~al\mbox{.}(2015b)]%
        {lin2015mscoco}
\bibfield{author}{\bibinfo{person}{Tsung-Yi Lin}, \bibinfo{person}{Michael Maire}, \bibinfo{person}{Serge Belongie}, \bibinfo{person}{Lubomir Bourdev}, \bibinfo{person}{Ross Girshick}, \bibinfo{person}{James Hays}, \bibinfo{person}{Pietro Perona}, \bibinfo{person}{Deva Ramanan}, \bibinfo{person}{C.~Lawrence Zitnick}, {and} \bibinfo{person}{Piotr Dollár}.} \bibinfo{year}{2015}\natexlab{b}.
\newblock \bibinfo{title}{Microsoft COCO: Common Objects in Context}.
\newblock
\showeprint[arxiv]{1405.0312}~[cs.CV]
\urldef\tempurl%
\url{https://arxiv.org/abs/1405.0312}
\showURL{%
\tempurl}


\bibitem[Liu and Wang(2021)]%
        {liu2021t}
\bibfield{author}{\bibinfo{person}{Qimin Liu} {and} \bibinfo{person}{Lijuan Wang}.} \bibinfo{year}{2021}\natexlab{}.
\newblock \showarticletitle{t-Test and ANOVA for data with ceiling and/or floor effects}.
\newblock \bibinfo{journal}{\emph{Behavior Research Methods}} \bibinfo{volume}{53}, \bibinfo{number}{1} (\bibinfo{year}{2021}), \bibinfo{pages}{264--277}.
\newblock


\bibitem[Lystb{\ae}k et~al\mbox{.}(2022)]%
        {lystbaek2022gaze}
\bibfield{author}{\bibinfo{person}{Mathias~N Lystb{\ae}k}, \bibinfo{person}{Peter Rosenberg}, \bibinfo{person}{Ken Pfeuffer}, \bibinfo{person}{Jens~Emil Gr{\o}nb{\ae}k}, {and} \bibinfo{person}{Hans Gellersen}.} \bibinfo{year}{2022}\natexlab{}.
\newblock \showarticletitle{Gaze-hand alignment: Combining eye gaze and mid-air pointing for interacting with menus in augmented reality}.
\newblock \bibinfo{journal}{\emph{Proceedings of the ACM on Human-Computer Interaction}} \bibinfo{volume}{6}, \bibinfo{number}{ETRA} (\bibinfo{year}{2022}), \bibinfo{pages}{1--18}.
\newblock


\bibitem[Masnadi et~al\mbox{.}(2020)]%
        {masnadi2020vriassist}
\bibfield{author}{\bibinfo{person}{Sina Masnadi}, \bibinfo{person}{Brian Williamson}, \bibinfo{person}{Andr{\'e}s N~Vargas Gonz{\'a}lez}, {and} \bibinfo{person}{Joseph~J LaViola}.} \bibinfo{year}{2020}\natexlab{}.
\newblock \showarticletitle{Vriassist: An eye-tracked virtual reality low vision assistance tool}. In \bibinfo{booktitle}{\emph{2020 IEEE Conference on Virtual Reality and 3D User Interfaces Abstracts and Workshops (VRW)}}. IEEE, \bibinfo{pages}{808--809}.
\newblock


\bibitem[Maus et~al\mbox{.}(2020)]%
        {maus2020gaze}
\bibfield{author}{\bibinfo{person}{Natalie Maus}, \bibinfo{person}{Dalton Rutledge}, \bibinfo{person}{Sedeeq Al-Khazraji}, \bibinfo{person}{Reynold Bailey}, \bibinfo{person}{Cecilia~Ovesdotter Alm}, {and} \bibinfo{person}{Kristen Shinohara}.} \bibinfo{year}{2020}\natexlab{}.
\newblock \showarticletitle{Gaze-guided magnification for individuals with vision impairments}. In \bibinfo{booktitle}{\emph{Extended Abstracts of the 2020 CHI Conference on Human Factors in Computing Systems}}. \bibinfo{pages}{1--8}.
\newblock


\bibitem[Meta(2026)]%
        {metaWristbandsTechnology}
\bibfield{author}{\bibinfo{person}{Meta}.} \bibinfo{year}{2026}\natexlab{}.
\newblock \bibinfo{title}{EMG Wristbands and Technology | Meta --- meta.com}.
\newblock \bibinfo{howpublished}{\url{https://www.meta.com/emerging-tech/emg-wearable-technology/?srsltid=AfmBOoqpQdYtnEolu9uYJfW4yEj5pSvi9ODVydniiwHdnPH4tvboUVrP}}.
\newblock
\newblock
\shownote{[Accessed 26-03-2026]}.


\bibitem[{Meta Platforms, Inc.}(2024)]%
        {meta_unity_spatial_anchors_best_practices}
\bibfield{author}{\bibinfo{person}{{Meta Platforms, Inc.}}} \bibinfo{year}{2024}\natexlab{}.
\newblock \bibinfo{booktitle}{\emph{Unity Spatial Anchors Best Practices}}.
\newblock
\urldef\tempurl%
\url{https://developers.meta.com/horizon/documentation/unity/unity-spatial-anchors-best-practices/}
\showURL{%
\tempurl}
\newblock
\shownote{Accessed: 2026-01-25}.


\bibitem[Minakata et~al\mbox{.}(2019)]%
        {minakata2019pointing}
\bibfield{author}{\bibinfo{person}{Katsumi Minakata}, \bibinfo{person}{John~Paulin Hansen}, \bibinfo{person}{I~Scott MacKenzie}, \bibinfo{person}{Per B{\ae}kgaard}, {and} \bibinfo{person}{Vijay Rajanna}.} \bibinfo{year}{2019}\natexlab{}.
\newblock \showarticletitle{Pointing by gaze, head, and foot in a head-mounted display}. In \bibinfo{booktitle}{\emph{Proceedings of the 11th ACM symposium on eye tracking research \& applications}}. \bibinfo{pages}{1--9}.
\newblock


\bibitem[Mohan et~al\mbox{.}(2018)]%
        {mohan2018dualgaze}
\bibfield{author}{\bibinfo{person}{Pallavi Mohan}, \bibinfo{person}{Wooi~Boon Goh}, \bibinfo{person}{Chi-Wing Fu}, {and} \bibinfo{person}{Sai-Kit Yeung}.} \bibinfo{year}{2018}\natexlab{}.
\newblock \showarticletitle{DualGaze: Addressing the midas touch problem in gaze mediated VR interaction}. In \bibinfo{booktitle}{\emph{2018 IEEE International Symposium on Mixed and Augmented Reality Adjunct (ISMAR-Adjunct)}}. IEEE, \bibinfo{pages}{79--84}.
\newblock


\bibitem[Moshtael et~al\mbox{.}(2015)]%
        {moshtael2015high}
\bibfield{author}{\bibinfo{person}{Howard Moshtael}, \bibinfo{person}{Tariq Aslam}, \bibinfo{person}{Ian Underwood}, {and} \bibinfo{person}{Baljean Dhillon}.} \bibinfo{year}{2015}\natexlab{}.
\newblock \showarticletitle{High tech aids low vision: a review of image processing for the visually impaired}.
\newblock \bibinfo{journal}{\emph{Translational vision science \& technology}} \bibinfo{volume}{4}, \bibinfo{number}{4} (\bibinfo{year}{2015}), \bibinfo{pages}{6--6}.
\newblock


\bibitem[M{\"u}ller et~al\mbox{.}(2017)]%
        {muller2017remote}
\bibfield{author}{\bibinfo{person}{Jens M{\"u}ller}, \bibinfo{person}{Roman R{\"a}dle}, {and} \bibinfo{person}{Harald Reiterer}.} \bibinfo{year}{2017}\natexlab{}.
\newblock \showarticletitle{Remote collaboration with mixed reality displays: How shared virtual landmarks facilitate spatial referencing}. In \bibinfo{booktitle}{\emph{Proceedings of the 2017 CHI Conference on Human Factors in Computing Systems}}. \bibinfo{pages}{6481--6486}.
\newblock


\bibitem[Mutasim et~al\mbox{.}(2021)]%
        {mutasim2021pinch}
\bibfield{author}{\bibinfo{person}{Aunnoy~K Mutasim}, \bibinfo{person}{Anil~Ufuk Batmaz}, {and} \bibinfo{person}{Wolfgang Stuerzlinger}.} \bibinfo{year}{2021}\natexlab{}.
\newblock \showarticletitle{Pinch, click, or dwell: Comparing different selection techniques for eye-gaze-based pointing in virtual reality}. In \bibinfo{booktitle}{\emph{Acm symposium on eye tracking research and applications}}. \bibinfo{pages}{1--7}.
\newblock


\bibitem[Nakamori et~al\mbox{.}(1997)]%
        {nakamori1997blinking}
\bibfield{author}{\bibinfo{person}{Katsu Nakamori}, \bibinfo{person}{Mikiko Odawara}, \bibinfo{person}{Toshiaki Nakajima}, \bibinfo{person}{Taku Mizutani}, {and} \bibinfo{person}{Kazuo Tsubota}.} \bibinfo{year}{1997}\natexlab{}.
\newblock \showarticletitle{Blinking is controlled primarily by ocular surface conditions}.
\newblock \bibinfo{journal}{\emph{American journal of ophthalmology}} \bibinfo{volume}{124}, \bibinfo{number}{1} (\bibinfo{year}{1997}), \bibinfo{pages}{24--30}.
\newblock


\bibitem[Namnakani et~al\mbox{.}(2023)]%
        {namnakani2023comparing}
\bibfield{author}{\bibinfo{person}{Omar Namnakani}, \bibinfo{person}{Yasmeen Abdrabou}, \bibinfo{person}{Jonathan Grizou}, \bibinfo{person}{Augusto Esteves}, {and} \bibinfo{person}{Mohamed Khamis}.} \bibinfo{year}{2023}\natexlab{}.
\newblock \showarticletitle{Comparing dwell time, pursuits and gaze gestures for gaze interaction on handheld mobile devices}. In \bibinfo{booktitle}{\emph{Proceedings of the 2023 CHI conference on human factors in computing systems}}. \bibinfo{pages}{1--17}.
\newblock


\bibitem[{National Eye Institute}(2024)]%
        {NEI_LowVision}
\bibfield{author}{\bibinfo{person}{{National Eye Institute}}.} \bibinfo{year}{2024}\natexlab{}.
\newblock \bibinfo{booktitle}{\emph{Low Vision}}.
\newblock U.S. National Institutes of Health.
\newblock
\urldef\tempurl%
\url{https://www.nei.nih.gov/learn-about-eye-health/eye-conditions-and-diseases/low-vision}
\showURL{%
\tempurl}
\newblock
\shownote{Accessed: 2025-11-02}.


\bibitem[Neuhold et~al\mbox{.}(2017)]%
        {neuhold2017mapillary}
\bibfield{author}{\bibinfo{person}{Gerhard Neuhold}, \bibinfo{person}{Tobias Ollmann}, \bibinfo{person}{Samuel Rota~Bulo}, {and} \bibinfo{person}{Peter Kontschieder}.} \bibinfo{year}{2017}\natexlab{}.
\newblock \showarticletitle{The mapillary vistas dataset for semantic understanding of street scenes}. In \bibinfo{booktitle}{\emph{Proceedings of the IEEE international conference on computer vision}}. \bibinfo{pages}{4990--4999}.
\newblock


\bibitem[Nguyen et~al\mbox{.}(2023)]%
        {nguyen2023hand}
\bibfield{author}{\bibinfo{person}{Richard Nguyen}, \bibinfo{person}{Charles Gouin-Vallerand}, {and} \bibinfo{person}{Maryam Amiri}.} \bibinfo{year}{2023}\natexlab{}.
\newblock \showarticletitle{Hand interaction designs in mixed and augmented reality head mounted display: a scoping review and classification}.
\newblock \bibinfo{journal}{\emph{Frontiers in Virtual Reality}}  \bibinfo{volume}{4} (\bibinfo{year}{2023}), \bibinfo{pages}{1171230}.
\newblock


\bibitem[NILSSON et~al\mbox{.}(1998)]%
        {nilsson1998location}
\bibfield{author}{\bibinfo{person}{ULLA~L NILSSON}, \bibinfo{person}{CHRISTINA FRENNESSON}, {and} \bibinfo{person}{SVEN ERIK~G NILSSON}.} \bibinfo{year}{1998}\natexlab{}.
\newblock \showarticletitle{Location and stability of a newly established eccentric retinal locus suitable for reading, achieved through training of patients with a dense central scotoma}.
\newblock \bibinfo{journal}{\emph{Optometry and Vision Science}} \bibinfo{volume}{75}, \bibinfo{number}{12} (\bibinfo{year}{1998}), \bibinfo{pages}{873--878}.
\newblock


\bibitem[Oflaz et~al\mbox{.}(2022)]%
        {oflaz2022short}
\bibfield{author}{\bibinfo{person}{Ay{\c{s}}e~Bozkurt Oflaz}, \bibinfo{person}{Banu~Turgut {\"O}zt{\"u}rk}, \bibinfo{person}{{\c{S}}aban G{\"o}n{\"u}l}, \bibinfo{person}{Berker Bakbak}, \bibinfo{person}{{\c{S}}ansal Gedik}, {and} \bibinfo{person}{S{\"u}leyman Okudan}.} \bibinfo{year}{2022}\natexlab{}.
\newblock \showarticletitle{Short-term clinical results of preferred retinal locus training}.
\newblock \bibinfo{journal}{\emph{Turk J Ophthalmol}} \bibinfo{volume}{52}, \bibinfo{number}{1} (\bibinfo{year}{2022}), \bibinfo{pages}{14--22}.
\newblock


\bibitem[Pal et~al\mbox{.}(2017)]%
        {pal2017agency}
\bibfield{author}{\bibinfo{person}{Joyojeet Pal}, \bibinfo{person}{Anandhi Viswanathan}, \bibinfo{person}{Priyank Chandra}, \bibinfo{person}{Anisha Nazareth}, \bibinfo{person}{Vaishnav Kameswaran}, \bibinfo{person}{Hariharan Subramonyam}, \bibinfo{person}{Aditya Johri}, \bibinfo{person}{Mark~S Ackerman}, {and} \bibinfo{person}{Sile O'Modhrain}.} \bibinfo{year}{2017}\natexlab{}.
\newblock \showarticletitle{Agency in assistive technology adoption: visual impairment and smartphone use in Bangalore}. In \bibinfo{booktitle}{\emph{Proceedings of the 2017 CHI conference on human factors in computing systems}}. \bibinfo{pages}{5929--5940}.
\newblock


\bibitem[Paulus and Remijn(2021)]%
        {paulus2021usability}
\bibfield{author}{\bibinfo{person}{Yesaya~Tommy Paulus} {and} \bibinfo{person}{Gerard~Bastiaan Remijn}.} \bibinfo{year}{2021}\natexlab{}.
\newblock \showarticletitle{Usability of various dwell times for eye-gaze-based object selection with eye tracking}.
\newblock \bibinfo{journal}{\emph{Displays}}  \bibinfo{volume}{67} (\bibinfo{year}{2021}), \bibinfo{pages}{101997}.
\newblock


\bibitem[Pozzo et~al\mbox{.}(2002)]%
        {pozzo2002coordination}
\bibfield{author}{\bibinfo{person}{Thierry Pozzo}, \bibinfo{person}{Paul~J Stapley}, {and} \bibinfo{person}{Charalambos Papaxanthis}.} \bibinfo{year}{2002}\natexlab{}.
\newblock \showarticletitle{Coordination between equilibrium and hand trajectories during whole body pointing movements}.
\newblock \bibinfo{journal}{\emph{Experimental Brain Research}} \bibinfo{volume}{144}, \bibinfo{number}{3} (\bibinfo{year}{2002}), \bibinfo{pages}{343--350}.
\newblock


\bibitem[{Pupil Labs}(2024)]%
        {PupilLabsFixationDetector2024}
\bibfield{author}{\bibinfo{person}{{Pupil Labs}}.} \bibinfo{year}{2024}\natexlab{}.
\newblock \bibinfo{title}{Pupil Labs Fixation Detector: Algorithm Description}.
\newblock
\urldef\tempurl%
\url{https://docs.google.com/document/d/1CZnjyg4P83QSkfHi_bjwSceWCTWvlVtbGWtuyajv5Jc/export?format=pdf}
\showURL{%
\tempurl}
\newblock
\shownote{Accessed: 2026-02-10}.


\bibitem[Qian and Teather(2017)]%
        {qian2017eyes}
\bibfield{author}{\bibinfo{person}{Yuan~Yuan Qian} {and} \bibinfo{person}{Robert~J Teather}.} \bibinfo{year}{2017}\natexlab{}.
\newblock \showarticletitle{The eyes don't have it: an empirical comparison of head-based and eye-based selection in virtual reality}. In \bibinfo{booktitle}{\emph{Proceedings of the 5th symposium on spatial user interaction}}. \bibinfo{pages}{91--98}.
\newblock


\bibitem[Radford et~al\mbox{.}(2022)]%
        {radford2022whisper}
\bibfield{author}{\bibinfo{person}{Alec Radford}, \bibinfo{person}{Jong~Wook Kim}, \bibinfo{person}{Tao Xu}, \bibinfo{person}{Greg Brockman}, \bibinfo{person}{Christine McLeavey}, {and} \bibinfo{person}{Ilya Sutskever}.} \bibinfo{year}{2022}\natexlab{}.
\newblock \bibinfo{title}{Whisper}.
\newblock \bibinfo{howpublished}{\url{https://github.com/openai/whisper}}.
\newblock
\newblock
\shownote{OpenAI, Automatic Speech Recognition System}.


\bibitem[Rahman et~al\mbox{.}(2015)]%
        {rahman2015corneal}
\bibfield{author}{\bibinfo{person}{Effie~Z Rahman}, \bibinfo{person}{Peter~K Lam}, \bibinfo{person}{Chia-Kai Chu}, \bibinfo{person}{Quianta Moore}, {and} \bibinfo{person}{Stephen~C Pflugfelder}.} \bibinfo{year}{2015}\natexlab{}.
\newblock \showarticletitle{Corneal sensitivity in tear dysfunction and its correlation with clinical parameters and blink rate}.
\newblock \bibinfo{journal}{\emph{American journal of ophthalmology}} \bibinfo{volume}{160}, \bibinfo{number}{5} (\bibinfo{year}{2015}), \bibinfo{pages}{858--866}.
\newblock


\bibitem[Richardson(2018)]%
        {richardson2018use}
\bibfield{author}{\bibinfo{person}{John~TE Richardson}.} \bibinfo{year}{2018}\natexlab{}.
\newblock \showarticletitle{The use of Latin-square designs in educational and psychological research}.
\newblock \bibinfo{journal}{\emph{Educational Research Review}}  \bibinfo{volume}{24} (\bibinfo{year}{2018}), \bibinfo{pages}{84--97}.
\newblock


\bibitem[Shi et~al\mbox{.}(2023)]%
        {shi2023exploring}
\bibfield{author}{\bibinfo{person}{Rongkai Shi}, \bibinfo{person}{Yushi Wei}, \bibinfo{person}{Xueying Qin}, \bibinfo{person}{Pan Hui}, {and} \bibinfo{person}{Hai-Ning Liang}.} \bibinfo{year}{2023}\natexlab{}.
\newblock \showarticletitle{Exploring gaze-assisted and hand-based region selection in augmented reality}.
\newblock \bibinfo{journal}{\emph{Proceedings of the ACM on Human-Computer Interaction}} \bibinfo{volume}{7}, \bibinfo{number}{ETRA} (\bibinfo{year}{2023}), \bibinfo{pages}{1--19}.
\newblock


\bibitem[Sidenmark et~al\mbox{.}(2024)]%
        {sidenmark2024cone}
\bibfield{author}{\bibinfo{person}{Ludwig Sidenmark}, \bibinfo{person}{Zibo Sun}, {and} \bibinfo{person}{Hans Gellersen}.} \bibinfo{year}{2024}\natexlab{}.
\newblock \showarticletitle{Cone\&Bubble: Evaluating Combinations of Gaze, Head and Hand Pointing for Target Selection in Dense 3D Environments}. In \bibinfo{booktitle}{\emph{2024 IEEE Conference on Virtual Reality and 3D User Interfaces Abstracts and Workshops (VRW)}}. IEEE, \bibinfo{pages}{642--649}.
\newblock


\bibitem[Simpson(2017)]%
        {simpson2017mini}
\bibfield{author}{\bibinfo{person}{Michael~J Simpson}.} \bibinfo{year}{2017}\natexlab{}.
\newblock \showarticletitle{Mini-review: Far peripheral vision}.
\newblock \bibinfo{journal}{\emph{Vision research}}  \bibinfo{volume}{140} (\bibinfo{year}{2017}), \bibinfo{pages}{96--105}.
\newblock


\bibitem[Stearns et~al\mbox{.}(2017)]%
        {stearns2017augmented}
\bibfield{author}{\bibinfo{person}{Lee Stearns}, \bibinfo{person}{Victor DeSouza}, \bibinfo{person}{Jessica Yin}, \bibinfo{person}{Leah Findlater}, {and} \bibinfo{person}{Jon~E Froehlich}.} \bibinfo{year}{2017}\natexlab{}.
\newblock \showarticletitle{Augmented reality magnification for low vision users with the microsoft hololens and a finger-worn camera}. In \bibinfo{booktitle}{\emph{Proceedings of the 19th International ACM SIGACCESS Conference on Computers and Accessibility}}. \bibinfo{pages}{361--362}.
\newblock


\bibitem[Stearns et~al\mbox{.}(2018)]%
        {stearns2018design}
\bibfield{author}{\bibinfo{person}{Lee Stearns}, \bibinfo{person}{Leah Findlater}, {and} \bibinfo{person}{Jon~E Froehlich}.} \bibinfo{year}{2018}\natexlab{}.
\newblock \showarticletitle{Design of an augmented reality magnification aid for low vision users}. In \bibinfo{booktitle}{\emph{Proceedings of the 20th international ACM SIGACCESS conference on computers and accessibility}}. \bibinfo{pages}{28--39}.
\newblock


\bibitem[Truong et~al\mbox{.}(2018)]%
        {truong2018capband}
\bibfield{author}{\bibinfo{person}{Hoang Truong}, \bibinfo{person}{Shuo Zhang}, \bibinfo{person}{Ufuk Muncuk}, \bibinfo{person}{Phuc Nguyen}, \bibinfo{person}{Nam Bui}, \bibinfo{person}{Anh Nguyen}, \bibinfo{person}{Qin Lv}, \bibinfo{person}{Kaushik Chowdhury}, \bibinfo{person}{Thang Dinh}, {and} \bibinfo{person}{Tam Vu}.} \bibinfo{year}{2018}\natexlab{}.
\newblock \showarticletitle{Capband: Battery-free successive capacitance sensing wristband for hand gesture recognition}. In \bibinfo{booktitle}{\emph{Proceedings of the 16th ACM Conference on Embedded Networked Sensor Systems}}. \bibinfo{pages}{54--67}.
\newblock


\bibitem[Tsandilas and Casiez(2024)]%
        {tsandilas2024illusory}
\bibfield{author}{\bibinfo{person}{Theophanis Tsandilas} {and} \bibinfo{person}{G{\'e}ry Casiez}.} \bibinfo{year}{2024}\natexlab{}.
\newblock \bibinfo{title}{The illusory promise of the Aligned Rank Transform—A systematic study of rank transformations}.
\newblock


\bibitem[Velloso et~al\mbox{.}(2017)]%
        {velloso2017motion}
\bibfield{author}{\bibinfo{person}{Eduardo Velloso}, \bibinfo{person}{Marcus Carter}, \bibinfo{person}{Joshua Newn}, \bibinfo{person}{Augusto Esteves}, \bibinfo{person}{Christopher Clarke}, {and} \bibinfo{person}{Hans Gellersen}.} \bibinfo{year}{2017}\natexlab{}.
\newblock \showarticletitle{Motion correlation: Selecting objects by matching their movement}.
\newblock \bibinfo{journal}{\emph{ACM Transactions on Computer-Human Interaction (TOCHI)}} \bibinfo{volume}{24}, \bibinfo{number}{3} (\bibinfo{year}{2017}), \bibinfo{pages}{1--35}.
\newblock


\bibitem[Wagner et~al\mbox{.}(2024)]%
        {wagner2024gaze}
\bibfield{author}{\bibinfo{person}{Uta Wagner}, \bibinfo{person}{Matthias Albrecht}, \bibinfo{person}{Andreas~Asferg Jacobsen}, \bibinfo{person}{Haopeng Wang}, \bibinfo{person}{Hans Gellersen}, {and} \bibinfo{person}{Ken Pfeuffer}.} \bibinfo{year}{2024}\natexlab{}.
\newblock \showarticletitle{Gaze, wall, and racket: Combining gaze and hand-controlled plane for 3D selection in virtual reality}.
\newblock \bibinfo{journal}{\emph{Proceedings of the ACM on Human-Computer Interaction}} \bibinfo{volume}{8}, \bibinfo{number}{ISS} (\bibinfo{year}{2024}), \bibinfo{pages}{189--213}.
\newblock


\bibitem[Wagner et~al\mbox{.}(2023)]%
        {wagner2023fitts}
\bibfield{author}{\bibinfo{person}{Uta Wagner}, \bibinfo{person}{Mathias~N Lystb{\ae}k}, \bibinfo{person}{Pavel Manakhov}, \bibinfo{person}{Jens Emil~Sloth Gr{\o}nb{\ae}k}, \bibinfo{person}{Ken Pfeuffer}, {and} \bibinfo{person}{Hans Gellersen}.} \bibinfo{year}{2023}\natexlab{}.
\newblock \showarticletitle{A fitts’ law study of gaze-hand alignment for selection in 3d user interfaces}. In \bibinfo{booktitle}{\emph{Proceedings of the 2023 CHI Conference on Human Factors in Computing Systems}}. \bibinfo{pages}{1--15}.
\newblock


\bibitem[Wang et~al\mbox{.}(2008)]%
        {wang2008investigating}
\bibfield{author}{\bibinfo{person}{Lijuan Wang}, \bibinfo{person}{Zhiyong Zhang}, \bibinfo{person}{John~J McArdle}, {and} \bibinfo{person}{Timothy~A Salthouse}.} \bibinfo{year}{2008}\natexlab{}.
\newblock \showarticletitle{Investigating ceiling effects in longitudinal data analysis}.
\newblock \bibinfo{journal}{\emph{Multivariate behavioral research}} \bibinfo{volume}{43}, \bibinfo{number}{3} (\bibinfo{year}{2008}), \bibinfo{pages}{476--496}.
\newblock


\bibitem[Wang et~al\mbox{.}(2025)]%
        {wang2025characterizing}
\bibfield{author}{\bibinfo{person}{Ru Wang}, \bibinfo{person}{Ruijia Chen}, \bibinfo{person}{Anqiao~Erica Cai}, \bibinfo{person}{Zhiyuan Li}, \bibinfo{person}{Sanbrita Mondal}, {and} \bibinfo{person}{Yuhang Zhao}.} \bibinfo{year}{2025}\natexlab{}.
\newblock \showarticletitle{Characterizing Visual Intents for People with Low Vision through Eye Tracking}. In \bibinfo{booktitle}{\emph{Proceedings of the 27th International ACM SIGACCESS Conference on Computers and Accessibility}}. \bibinfo{pages}{1--18}.
\newblock


\bibitem[Wang et~al\mbox{.}(2024)]%
        {wang2024gazeprompt}
\bibfield{author}{\bibinfo{person}{Ru Wang}, \bibinfo{person}{Zach Potter}, \bibinfo{person}{Yun Ho}, \bibinfo{person}{Daniel Killough}, \bibinfo{person}{Linxiu Zeng}, \bibinfo{person}{Sanbrita Mondal}, {and} \bibinfo{person}{Yuhang Zhao}.} \bibinfo{year}{2024}\natexlab{}.
\newblock \showarticletitle{GazePrompt: Enhancing Low Vision People's Reading Experience with Gaze-Aware Augmentations}. In \bibinfo{booktitle}{\emph{Proceedings of the 2024 CHI Conference on Human Factors in Computing Systems}}. \bibinfo{pages}{1--17}.
\newblock


\bibitem[Wang et~al\mbox{.}(2023)]%
        {wang2023understanding}
\bibfield{author}{\bibinfo{person}{Ru Wang}, \bibinfo{person}{Linxiu Zeng}, \bibinfo{person}{Xinyong Zhang}, \bibinfo{person}{Sanbrita Mondal}, {and} \bibinfo{person}{Yuhang Zhao}.} \bibinfo{year}{2023}\natexlab{}.
\newblock \showarticletitle{Understanding how low vision people read using eye tracking}. In \bibinfo{booktitle}{\emph{Proceedings of the 2023 CHI Conference on Human Factors in Computing Systems}}. \bibinfo{pages}{1--17}.
\newblock


\bibitem[Watson et~al\mbox{.}(2006)]%
        {watson2006effects}
\bibfield{author}{\bibinfo{person}{Gale~R Watson}, \bibinfo{person}{Ronald~A Schuchard}, \bibinfo{person}{William~R De~l'Aune}, {and} \bibinfo{person}{Erica Watkins}.} \bibinfo{year}{2006}\natexlab{}.
\newblock \showarticletitle{Effects of preferred retinal locus placement on text navigation and development of advantageous trained retinal locus}.
\newblock \bibinfo{journal}{\emph{Journal of Rehabilitation Research and Development}} \bibinfo{volume}{43}, \bibinfo{number}{6} (\bibinfo{year}{2006}), \bibinfo{pages}{761}.
\newblock


\bibitem[Wei et~al\mbox{.}(2025)]%
        {wei2025reevaluating}
\bibfield{author}{\bibinfo{person}{Yushi Wei}, \bibinfo{person}{Rongkai Shi}, \bibinfo{person}{Sen Zhang}, \bibinfo{person}{Anil~Ufuk Batmaz}, \bibinfo{person}{Pan Hui}, {and} \bibinfo{person}{Hai-Ning Liang}.} \bibinfo{year}{2025}\natexlab{}.
\newblock \showarticletitle{Reevaluating the Gaze Cursor in Virtual Reality: A Comparative Analysis of Cursor Visibility, Confirmation Mechanisms, and Task Paradigms}.
\newblock \bibinfo{journal}{\emph{IEEE Transactions on Visualization and Computer Graphics}} (\bibinfo{year}{2025}).
\newblock


\bibitem[Whiffing et~al\mbox{.}(2026)]%
        {whiffing2026understanding}
\bibfield{author}{\bibinfo{person}{James Whiffing}, \bibinfo{person}{Tobias Langlotz}, \bibinfo{person}{Christof Lutteroth}, \bibinfo{person}{Adwait Sharma}, {and} \bibinfo{person}{Christopher Clarke}.} \bibinfo{year}{2026}\natexlab{}.
\newblock \showarticletitle{Understanding Freehand Cursorless Pointing Variability and Its Impact on Selection Performance}.
\newblock \bibinfo{journal}{\emph{ACM Transactions on Computer-Human Interaction}} \bibinfo{volume}{33}, \bibinfo{number}{1} (\bibinfo{year}{2026}), \bibinfo{pages}{1--47}.
\newblock


\bibitem[Whittaker et~al\mbox{.}(1988)]%
        {whittaker1988eccentric}
\bibfield{author}{\bibinfo{person}{Stephen~G Whittaker}, \bibinfo{person}{James Budd}, {and} \bibinfo{person}{RW Cummings}.} \bibinfo{year}{1988}\natexlab{}.
\newblock \showarticletitle{Eccentric fixation with macular scotoma.}
\newblock \bibinfo{journal}{\emph{Investigative ophthalmology \& visual science}} \bibinfo{volume}{29}, \bibinfo{number}{2} (\bibinfo{year}{1988}), \bibinfo{pages}{268--278}.
\newblock


\bibitem[Wobbrock et~al\mbox{.}(2011)]%
        {wobbrock2011aligned}
\bibfield{author}{\bibinfo{person}{Jacob~O Wobbrock}, \bibinfo{person}{Leah Findlater}, \bibinfo{person}{Darren Gergle}, {and} \bibinfo{person}{James~J Higgins}.} \bibinfo{year}{2011}\natexlab{}.
\newblock \showarticletitle{The aligned rank transform for nonparametric factorial analyses using only anova procedures}. In \bibinfo{booktitle}{\emph{Proceedings of the SIGCHI conference on human factors in computing systems}}. \bibinfo{pages}{143--146}.
\newblock


\bibitem[Xiao et~al\mbox{.}(2021)]%
        {xiao2021let}
\bibfield{author}{\bibinfo{person}{Ziang Xiao}, \bibinfo{person}{Sarah Mennicken}, \bibinfo{person}{Bernd Huber}, \bibinfo{person}{Adam Shonkoff}, {and} \bibinfo{person}{Jennifer Thom}.} \bibinfo{year}{2021}\natexlab{}.
\newblock \showarticletitle{Let me ask you this: How can a voice assistant elicit explicit user feedback?}
\newblock \bibinfo{journal}{\emph{Proceedings of the ACM on Human-Computer Interaction}} \bibinfo{volume}{5}, \bibinfo{number}{CSCW2} (\bibinfo{year}{2021}), \bibinfo{pages}{1--24}.
\newblock


\bibitem[Xu et~al\mbox{.}(2019)]%
        {xu2019pointing}
\bibfield{author}{\bibinfo{person}{Wenge Xu}, \bibinfo{person}{Hai-Ning Liang}, \bibinfo{person}{Anqi He}, {and} \bibinfo{person}{Zifan Wang}.} \bibinfo{year}{2019}\natexlab{}.
\newblock \showarticletitle{Pointing and selection methods for text entry in augmented reality head mounted displays}. In \bibinfo{booktitle}{\emph{2019 IEEE International Symposium on Mixed and Augmented Reality (ISMAR)}}. IEEE, \bibinfo{pages}{279--288}.
\newblock


\bibitem[Zhao et~al\mbox{.}(2020)]%
        {zhao2020effectiveness}
\bibfield{author}{\bibinfo{person}{Yuhang Zhao}, \bibinfo{person}{Elizabeth Kupferstein}, \bibinfo{person}{Hathaitorn Rojnirun}, \bibinfo{person}{Leah Findlater}, {and} \bibinfo{person}{Shiri Azenkot}.} \bibinfo{year}{2020}\natexlab{}.
\newblock \showarticletitle{The effectiveness of visual and audio wayfinding guidance on smartglasses for people with low vision}. In \bibinfo{booktitle}{\emph{Proceedings of the 2020 CHI conference on human factors in computing systems}}. \bibinfo{pages}{1--14}.
\newblock


\bibitem[Zhao et~al\mbox{.}(2015)]%
        {zhao2015foresee}
\bibfield{author}{\bibinfo{person}{Yuhang Zhao}, \bibinfo{person}{Sarit Szpiro}, {and} \bibinfo{person}{Shiri Azenkot}.} \bibinfo{year}{2015}\natexlab{}.
\newblock \showarticletitle{Foresee: A customizable head-mounted vision enhancement system for people with low vision}. In \bibinfo{booktitle}{\emph{Proceedings of the 17th international ACM SIGACCESS conference on computers \& accessibility}}. \bibinfo{pages}{239--249}.
\newblock


\bibitem[Zhao et~al\mbox{.}(2016)]%
        {zhao2016cuesee}
\bibfield{author}{\bibinfo{person}{Yuhang Zhao}, \bibinfo{person}{Sarit Szpiro}, \bibinfo{person}{Jonathan Knighten}, {and} \bibinfo{person}{Shiri Azenkot}.} \bibinfo{year}{2016}\natexlab{}.
\newblock \showarticletitle{CueSee: exploring visual cues for people with low vision to facilitate a visual search task}. In \bibinfo{booktitle}{\emph{Proceedings of the 2016 ACM International Joint Conference on Pervasive and Ubiquitous Computing}}. \bibinfo{pages}{73--84}.
\newblock


\bibitem[Zhao et~al\mbox{.}(2019)]%
        {zhao2019designing}
\bibfield{author}{\bibinfo{person}{Yuhang Zhao}, \bibinfo{person}{Sarit Szpiro}, \bibinfo{person}{Lei Shi}, {and} \bibinfo{person}{Shiri Azenkot}.} \bibinfo{year}{2019}\natexlab{}.
\newblock \showarticletitle{Designing and evaluating a customizable head-mounted vision enhancement system for people with low vision}.
\newblock \bibinfo{journal}{\emph{ACM Transactions on Accessible Computing (TACCESS)}} \bibinfo{volume}{12}, \bibinfo{number}{4} (\bibinfo{year}{2019}), \bibinfo{pages}{1--46}.
\newblock


\end{thebibliography}
